\documentclass[pdflatex,sn-mathphys-num]{sn-jnl}

\usepackage{graphicx}%
\usepackage{multirow}%
\usepackage{amsmath,amssymb,amsfonts}%
\usepackage{amsthm}%
\usepackage{mathrsfs}%
\DeclareMathAlphabet{\mathscr}{U}{rsfs}{m}{n}
\usepackage[title]{appendix}%
\usepackage{xcolor}%
\usepackage{textcomp}%
\usepackage{manyfoot}%
\usepackage{booktabs}%
\usepackage{algorithm}%
\usepackage{algorithmicx}%
\usepackage{algpseudocode}%
\usepackage{listings}%
\usepackage{longtable}

\usepackage{mathrsfs}

\usepackage{float}      

\usepackage{caption}

\theoremstyle{thmstyleone}%
\theoremstyle{thmstyletwo}%

\theoremstyle{thmstylethree}%

\begin{document}

\title[Article Title]{How to Detect Information Voids Using Longitudinal Data from Social Media and Web Searches}

\author[1]{\fnm{Irene} \sur{Scalco}}\email{irene.scalco@uniroma1.it}
\author[2]{\fnm{Francesco} \sur{Gesualdo}}\email{francesco.gesualdo@med.unipi.it}
\author[1]{\fnm{Roy} \sur{Cerqueti}}\email{roy.cerqueti@uniroma1.it}
\author*[3]{\fnm{Matteo} \sur{Cinelli}}\email{matteo.cinelli@uniroma1.it}

\affil[1]{\orgdiv{Department of Social Sciences and Economics}, \orgname{Sapienza University of Rome}, \orgaddress{\street{P.le Aldo Moro, 5}, \city{Rome}, \postcode{00185}, \country{Italy}}}

\affil[2]{\orgdiv{Department of Translational Research and New Technologies in Medicine and Surgery}, \orgname{University of Pisa}, \orgaddress{\street{Via Savi, 10}, \city{Pisa}, \postcode{56126}, \country{Italy}}}

\affil*[3]{\orgdiv{Department of Computer Science}, \orgname{Sapienza University of Rome}, \orgaddress{\street{Viale Regina Elena 295}, \city{Rome}, \postcode{00161}, \country{Italy}}}



\abstract{
The model of the attention economy, where content producers compete for the attention of users, relies on two key forces: information supply and demand. This study leverages the feedback loop between these forces to develop a method for detecting and quantifying information voids, i.e., periods in which little or no reliable information is available on a given topic. 
Using a case study on COVID-19 vaccines rollout in six European countries, and drawing on data from multiple platforms including Facebook, Google, Twitter, Wikipedia, and online news outlets, we examine how information voids emerge, persist and correlate with a decline in the proportion of high-quality information circulating online. By conceptualising information voids as a specific regime of information spreading, we also quantify their counterpart, information overabundance, which constitute a central component of the current definition of infodemic. 
We show that information voids are associated with a higher prevalence of misinformation, thus representing problematic hotspots in which individuals are more likely to be misled by low-quality online content. Overall, our findings provide empirical support for the inclusion of information voids in mechanistic explanations of misinformation emergence.
}


\maketitle

\section*{Introduction}\label{sec1}

Social media platforms have profoundly reshaped how individuals obtain information and form opinions \cite{schmidt2017anatomy}. Once primarily channels for entertainment and interpersonal communication, they now serve as major gateways to news and public information \cite{avalle2024persistent, aldayel2021stance, cinelli2021echo, tucker2018social}.   
Within these platforms, and more in general on the internet, the widespread dissemination of misinformation represents one of the central, contemporary challenges \cite{globalrisks2024, lazer2018science, del2016spreading}. Social media amplifies false or misleading content across diverse domains \cite{van2022misinformation, pennycook2021psychology, allcott2019trends, van2017inoculating} yet the full range of factors driving its spread remains incompletely understood, including the impact of unreliable sources \cite{bovet2019influence, chou2018addressing, vosoughi2018spread}, rapid virality \cite{bak2022combining,sangiorgio2025evaluating}, and the overwhelming diversity of online content \cite{valensise2021entropy}.

More recently, the concept of information voids has become a focal point in this debate \cite{dataandsociety2019datavoids, purnat2021digital, purnat2021infodemic, nature2024_infovoids, aslett2024online}. Information voids refer to information spaces that lack a sufficient share of reliable, evidence-based content, particularly on emerging or rapidly developing topics of public interest. These voids pose a significant threat to the quality of public discourse: in the absence of authoritative and reliable sources, users searching for information may encounter vague, misleading, or outright false content. Consequently, information voids create fertile conditions for the diffusion of misinformation \cite{nature2024_infovoids, aslett2024online}, especially during periods of high public interest or collective uncertainty. 

Despite increasing concern about the risks posed by information voids, neither consolidated methods to measure them, nor means for building a solid ground truth for their presence exist. In fact, evidence of information voids typically relies on qualitative assessments or institutional warnings issued during the period in which they occur \cite{WHO2024FalseInformationToolkit}. This lack of operational definitions and validated benchmarks highlights a critical gap in experts ability to systematically detect and analyse the presence and the risks associated to information voids.

To address this gap, we propose a quantitative method, based on the demand and supply of information, that enables a systematic analysis of information dynamics, including voids and overabundance of available content, over time and across topics. In more detail, the proposed method detects how structural imbalances between information demand and supply can lead to distinct information states, such as voids -- where demand significantly exceeds supply -- or overabundance -- where supply overwhelms demand. In this study, information supply is defined as the volume of content produced on social media platforms (Facebook, Twitter) and online news outlets, measured using GDELT data \cite{gdelt_summary_api} while information demand is defined as online user searches, measured through Google Trends data \cite{nghiem2016analysis} and Wikipedia page views \cite{wikimedia_pageviews}.

Specifically, our method relies on a data science pipeline based on time series analysis that includes series rescaling and comparison of heterogeneous information sources, the definition of the information gap, the detection of anomalies in the gap time series, and the classification of the system into regimes of information diffusion, with information overabundance and information voids representing its extremes.
Our pipeline is first tested on synthetic data and then evaluated on a case study that has been described, based on qualitative assessment, as a potential instance of information voids \cite{WHO2025SocialListening, purnat2021digital, piltch2023were}. In more detail, we focus on six countries - Denmark, France, Germany, Italy, Spain, United Kingdom - over the period January 1, 2020 to April 30, 2021, a time frame that captures both the initial phase of the COVID-19 pandemic and the early stages of vaccine roll-out. During this time window, social media platforms and online information played a fundamental role in shaping the public response to the pandemic \cite{cinelli2020covid, tangcharoensathien2020framework, gallotti2020assessing, briand2021infodemics,loomba2021measuring,cinelli2025infodemic}, serving as key channels of communication between governments, health institutions, and citizens. 

The results of our method shows that sensitive moments (such as the vaccine roll-out, the EMA’s authorization of the Moderna vaccine, and the temporary suspension of the AstraZeneca vaccine) lead to an immediate imbalance between information demand and supply, indicating a marked responsiveness of the information ecosystem to these events. During those moments, the method detected several information voids that, beyond showing a high persistence over time (in terms of consecutive days), were associated with an increasing likelihood of user exposure to lower-quality content, including misinformation.

The proposed approach is both replicable and adaptable across different domains, making it suitable for analysing a wide range of informational contexts, including those related to public health, politics, environmental issues, and emerging crises.

The paper is organized as follows. Data science pipeline section describes the analytical framework of the study, including the quantification of information supply and demand, the rescaling of the corresponding time series, the definition of the information gap, called delta, and the detection of anomalies. Results and discussion section reports the application of the pipeline to real-world data using targeted case studies and examines the relationship between the presence of information voids and the reduction in the overall quality of information circulating online. Finally, a Methods section provides details on data collection and other methods used in the paper.

\section*{Data Science Pipeline}\label{sec2}

Modelling information voids as imbalances between information demand and supply requires quantifying both the volume of available content (supply) and the corresponding information needs (demand). A high level of supply relative to demand characterises an information ecosystem marked by information overabundance, whereas low supply combined with high demand indicates a gap in available information. However, it is the magnitude of this gap that determines the presence of an information void, thus requiring suitable methods to detect anomalous imbalances between supply and demand. Finally, the presence of information voids signals a structural imbalance and, in specific situations, these voids may become hotspots for misinformation. In fact, misinformation is typically characterized by a faster cycle of production and diffusion~\cite{vosoughi2018spread} compared to reliable information, which requires longer production times due to scientific validation or technical verification.

\subsection*{Quantifying information supply and demand}
The online information ecosystem can be conceptualised as a market shaped by two fundamental forces: the demand and supply of information \cite{gravino2022supply, gravino2024online}. Information demand reflects the topics and issues that people actively seek out or express interest in, what individuals want or need to know. Conversely, information supply refers to the volume of content circulating online across various channels, including digital news outlets and social media platforms. 
To operationalise these concepts, we adopt proxies for both demand and supply. For information demand, we rely on two complementary data sources: Wikipedia page-views and Google Trends. These platforms collectively capture patterns of public attention and interest in a timely and quantifiable manner. Wikipedia page-views provide daily-level data, yielding a time series that reflects the number of visits to specific pages over time. Prior research supports the use of Wikipedia traffic as a reliable indicator of public interest. A previous study \cite{yoshida2015wikipedia} demonstrated that, frequently, web-searched keywords exhibit strong correlations with Wikipedia page-views, underscoring their validity as a proxy for collective information-seeking behaviour. Beyond web search trends, Wikipedia traffic data have been successfully applied in diverse domains - from financial market analysis \cite{moat2013quantifying} to public health surveillance and policy research \cite{alibudbud2023wikipedia}.
Similarly, Google Trends is one of the most widely used tools in the socio-economic \cite{cerqueti2024anxiety} and behavioural sciences \cite{cerqueti2024investors, cerqueti2024portfolio, cebrian2023google, costola2021google}, and has proven effective in capturing patterns of searching behaviour across diverse contexts \cite{jun2018ten, zhang2018using}. By providing access to anonymised and aggregated search queries from the Google Search Engine, the platform offers valuable insights into the dynamics of public interest across regions and over time. 

On the supply side, we estimate the volume of topical information produced and circulating online. Following established approaches, information supply is represented by the volume of posts on platforms \cite{scalco2025modelling, gravino2024online, gravino2022supply} like Facebook and Twitter (now X) and articles published online, as recorded in the GDELT Summary database \cite{gdelt_summary_api}. The GDELT Summary database aggregates and summarises global news media coverage by reporting the volume of published articles related to selected topics.

In this context, we implemented a workflow pipeline designed to detect and characterise the dynamics of the information ecosystem, with the primary objective of identifying periods of information overabundance, when supply significantly exceeds demand, and information voids, where public interest emerges in the absence of sufficient content supply. 

\subsubsection*{Rescaling and harmonising supply and demand data} \label{sec:Rescaling and harmonising supply and demand}

A key methodological challenge in comparing and jointly using information demand and supply lies in the heterogeneity of the underlying data sources, which may differ in a number of aspects, including scale, temporal resolution, and content generation mechanisms. To ensure meaningful cross-source comparisons, all time series have to be rescaled to a common reference frame, where each observation represents the relative volume of information requested or supplied at a given time. 

Specifically, assume that each series is observed $T$ consecutive times and label the generic series as $x=(x_1,\dots, x_T)$. Time $t$ represents a specified time periodicity (e.g., days, weeks). In our model, $x$ may represent either supply $s$ or demand $d$. Each element of $x$ is rescaled by its expected value over the full observation period. 
Formally, the rescaled value $\hat{x}_{t}$ of a data point $x_{t}$ is defined as:
\begin{equation}
    \hat{x}_{t} = \frac{x_{t}}{\mathbb{E}(x)}
\end{equation}
where $\mathbb{E}(x)$ is the expected value of $x$, and it is computed as the arithmetic mean of the components of $x$. This rescaling procedure allows comparability of the considered series. In addition, rescaling by the expected value emphasizes deviations from typical behaviour, making it possible to identify unusually high or low activity even when observations span several orders of magnitude. This is particularly relevant in an online information ecosystem, where activity distributions are often highly skewed.

\subsubsection*{Constructing the information delta measure}
To quantify the relationship between information supply and demand over time, we fix at the time $t$ and introduce 
\begin{equation}
    \delta_t = \hat{s}_t - \hat{d}_t
    \label{eq_delta}
\end{equation}
where $\hat{s}_t$ and $\hat{d}_t$ represent the rescaled values of information supply and demand at time $t$, respectively. 
This measure captures the instantaneous imbalance between the two forces that shape the information ecosystem. Positive values of $\delta_t$ indicate periods in which information supply exceeds information demand, while negative values denote moments of information scarcity, when demand exceeds supply. We denote the quantity described by the $\delta_t$ simply as \textit{information delta}. 

\subsubsection*{Detect anomalies in the delta series}
While the information delta provides a useful first-level characterisation of supply–demand imbalances, further refinement is required to distinguish ordinary fluctuations from statistically meaningful deviations. A straightforward approach would be to apply a fixed linear threshold, for example flagging values that deviate by more than three times the expected level as indicative of information overload or information voids. However, such thresholds are often overly simplistic and highly sensitive to short-term variability.

We therefore adopt a more robust strategy based on time-series anomaly detection using conservative statistical criteria, which enables a more reliable identification of exceptional imbalances between information supply and demand. Specifically, we apply an anomaly detection procedure to the information delta series in order to detect statistically significant departures from the expected balance. This approach not only identifies anomalous periods, but also estimates tolerance bands around the equilibrium point ($\delta_t = 0$), within which the information ecosystem can be considered stable. Values exceeding these bands signal potential periods of systemic stress, during which the balance between content production and public attention may temporarily break down.

Anomalies are defined as observations that deviate significantly, according to a given threshold, from the overall behaviour of the data \cite{liu2023anomaly, schmidl2022anomaly, teng2010anomaly}, to the extent that they may plausibly originate from a distinct underlying process \cite{hawkins1980identification}. Operationally, a data point is classified as anomalous when it diverges from expected temporal patterns or differs substantially from historical observations \cite{blazquez2021review}. This approach goes beyond a purely descriptive inspection of peaks and troughs by introducing a systematic, transparent, and replicable procedure for identifying exceptional information events. 

The method relies on the distributional properties of the time series after isolating its structural components. The original series is first decomposed using Seasonal–Trend decomposition, separating it into trend, seasonal and remainder components. Anomaly detection is then applied exclusively to the remainder component, which captures irregular, transient fluctuations not explained by the underlying trend or seasonal structure and is therefore most informative for identifying unusual or unexpected events. Outliers in the remainder component are identified using the interquartile range (IQR) criterion, whereby observations falling outside the interval $[Q_1 - k \cdot \mathrm{IQR},\; Q_3 + k \cdot \mathrm{IQR}]$ are classified as anomalies, with $Q_1$ and $Q_3$ denoting the first and third quartiles of the $\delta_t$ distribution. More details on the anomaly detection algorithm employed can be found in Section \ref{Anomaly Detection}. 

\subsection*{Summary of the pipeline for overabundance and voids detection}
\begin{figure}[H]
    \centering
    \includegraphics[width=1\linewidth, trim=3.5cm 7cm 3.5cm 7cm, clip]{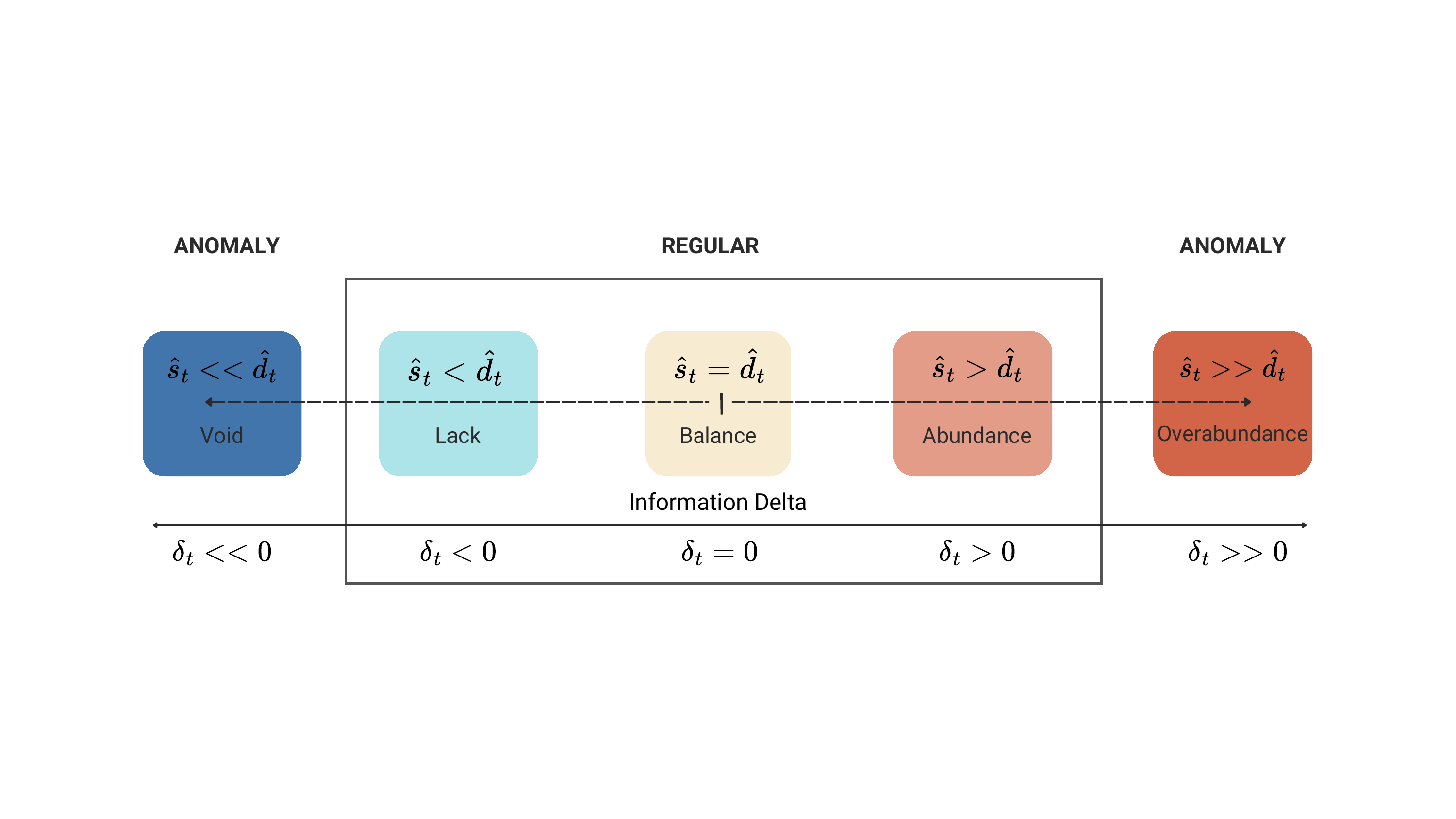}
    \caption{Definition of the information delta regimes.}
    \label{fig:delta_states}
\end{figure}

This pipeline enables the analysis and characterization of the online information ecosystem using longitudinal data from social media and web searches.  
As shown in Figure~\ref{fig:delta_states}, we develop a classification into regimes consisting of five distinct states derived from the output of the delta anomaly detection: Void, Lack, Balance, Abundance, and Overabundance. 
These five states are grouped into two broader categories that distinguish between two macro-states: the regular state and the anomaly state. The most critical conditions within the anomaly state correspond to voids and overabundance. A void arises when information demand exceeds supply, resulting in strongly negative delta values; this can occur even when supply is high, as long as demand remain significantly higher. Conversely, overabundance refers to situations in which content production is extremely high relative to low information demand, resulting in delta values well above zero. The "regular state" (including Lack, Balance, and Abundance) comprehends all, non-anomalous situations in which information supply and demand are broadly aligned or where their imbalance remains within tolerance bounds and does not qualify as an anomaly.
Algorithm~\ref{alg:pipeline} summarizes the pipeline  for the detection of overabundance and void periods.
\begin{algorithm}[H]
\caption{Detection of Overabundance and Void Periods}
\label{alg:pipeline}
\begin{algorithmic}[1]
    \State \textbf{Input:} Time series of information supply $s_t$ and demand $d_t$, anomaly detection method.
    \State \textbf{Output:} Labels $\mathcal{L}_t \in \{\text{Overabundance}, \text{Regular}, \text{Void}\}$,
    
    \State Rescale both series by their respective expectations:
    \[
    \hat{s}_t = \frac{s_t}{\mathbb{E}[s_t]}, \quad
    \hat{d}_t = \frac{d_t}{\mathbb{E}[d_t]}
    \]
    \State Compute the delta time series:
    \[
    \delta_t = \hat{s}_t - \hat{d}_t
    \]
    \State Detect anomalies in $\delta_t$ using the Anomaly Detection Method
    \For{each detected anomaly at time $t$}
        \If{$\delta_t > 0$}
            \State Label as Overabundance
        \ElsIf{$\delta_t < 0$}
            \State Label as Void
        \Else
            \State Label as Regular
        \EndIf
    \EndFor
    \State \textbf{Return} set of labelled data points $\mathcal{L}_t$
\end{algorithmic}
\end{algorithm}
\subsection*{Pipeline evaluation using synthetic data} 
\begin{figure}[H]
    \centering
    \includegraphics[width=1\linewidth, trim=0cm 7cm 0cm 7cm, clip]{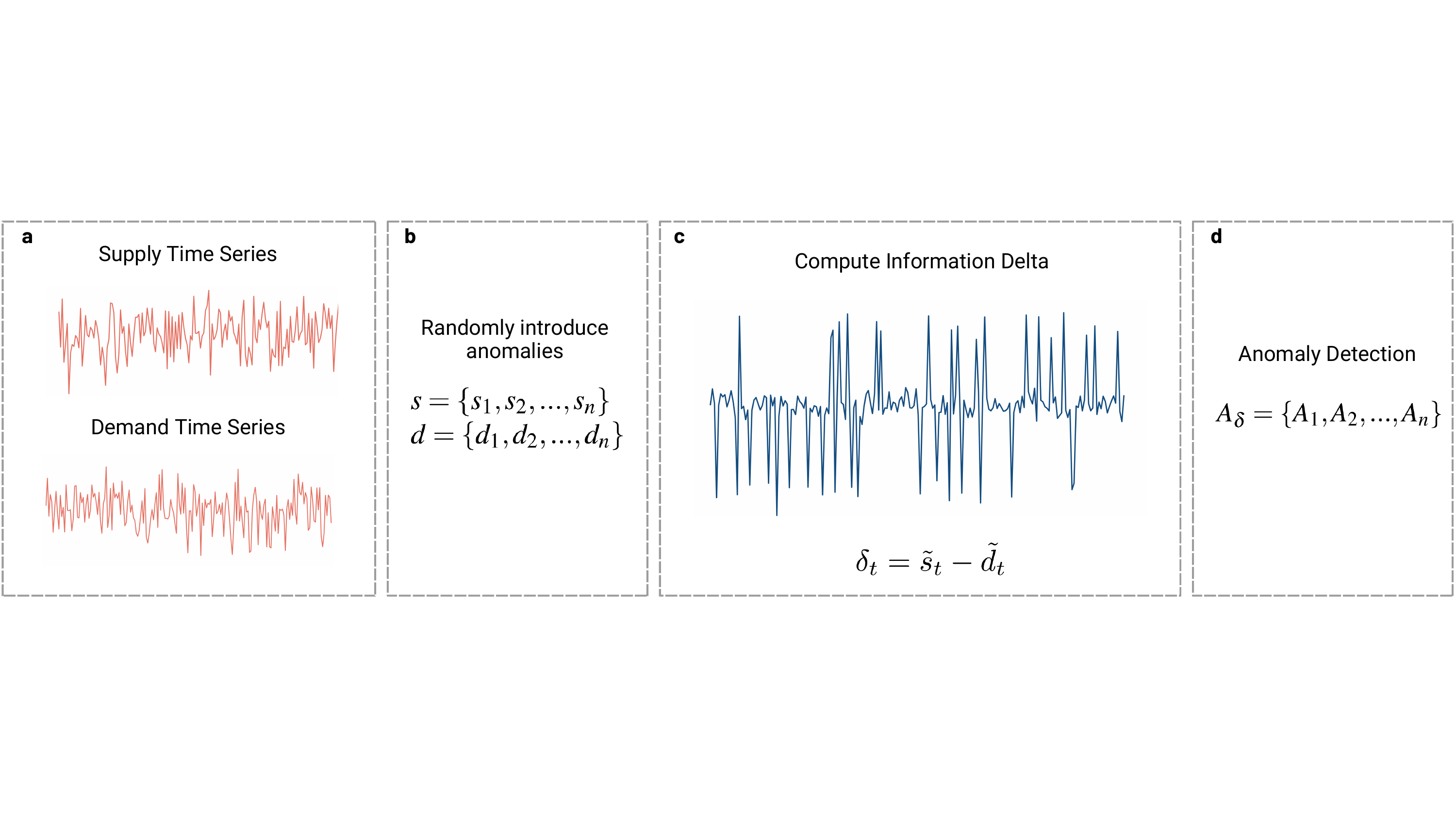}
    \caption{Pictorial representation of the synthetic data simulation process. a) Supply and demand are generated from Gaussian distributions. b) Anomalies of varying magnitudes are randomly added to the series, generating artificial spikes. c) The information delta is computed. d) The anomaly detection algorithm is applied to the resulting delta time series.}
    \label{fig:synt_data_exp}
\end{figure}

To assess the methodological robustness of the proposed framework and evaluate its applicability across different contexts, we first apply the pipeline using synthetic data. Synthetic time series representing information supply and demand are generated from Gaussian distributions $N(\mu,\sigma^2)$ with $(\mu = 10,\sigma^2 = 1)$ and tweaked to approximate the statistical properties of empirical data. In particular, the mean and standard deviation were selected so as to keep low the probability of the occurrence of negative values in the series; any residual negative instances were subsequently clipped to zero.
Furthermore, to replicate extreme deviations observed in real-world conditions, anomalies of varying magnitudes, ranging from +1 to +15 standard deviations, were randomly added into the series, thus generating artificial spikes. The upper bound of $15 \sigma$ was chosen because the maximum $Z-score$ observed in the empirical data was exceptionally high, which justified the inclusion of large anomalies in the simulation (further details are provided in the SI). For each anomaly value $\sigma \in \{1,1.5,...,15\}$, 20 random data points were perturbed of a $\sigma$ value in either the positive or negative direction, and each simulation (one for each $\sigma$ value) was repeated ten times to ensure statistical robustness. After generating the anomalies, the same analytical pipeline applied to the empirical data was used: the series were rescaled, the information delta (i.e., the difference between rescaled supply and demand) was computed, and the anomaly detection algorithm was applied to the resulting delta time series. The data generation and pipeline applied to synthetic data is illustrated in Figure~\ref{fig:synt_data_exp}. 
The results indicate that the model reliably detects higher-magnitude anomalies, corresponding to extreme imbalances between information demand and supply. As shown in Table~\ref{tab:synth_data}, performance metrics, including precision (proportion of detected anomalies that are truly correct) and F1 score (overall assessment of the model's accuracy), indicate that the model achieves a mean precision higher than 90\% for anomaly levels above $6\sigma$ and a mean F1 score higher than 0.68 for anomaly levels above $9\sigma$.
\begin{table}[hpt]
    \centering
    \begin{tabular}{c c c}
        \hline
        $\sigma$ & Mean Precision & Mean F1 Score \\
        \hline
        1 & 0 & 0 \\
        2 & 0 & 0 \\
        3 & 0.1 & 0.005 \\
        4 & 0.2 & 0.021 \\
        5 & 0.5 & 0.067 \\
        6 & 0.9 & 0.137 \\
        7 & 1.0 & 0.384 \\
        8 & 1.0 & 0.604 \\
        9 & 1.0 & 0.682 \\
        10 & 1.0 & 0.689 \\
        11 & 1.0 & 0.689 \\
        12 & 1.0 & 0.691 \\
        13 & 1.0 & 0.686 \\
        14 & 1.0 & 0.682 \\
        15 & 1.0 & 0.693 \\
        \hline
    \end{tabular}
    \captionsetup{width=\textwidth}
    \caption{Average Precision and F1 score as a function of $\sigma$.}
    \label{tab:synth_data}
\end{table}
Additional details are provided in the SI, which includes a table reporting all precision and F1 Score values, together with a description of how they are computed.

\section*{Results and Discussion} 
\subsubsection*{Information delta}
To validate the proposed pipeline, we applied it to real-world data collected between January 2020 and April 2021, focusing on the public discourse surrounding the five major COVID-19 vaccine brands (AstraZeneca, Moderna, Johnson \& Johnson, Pfizer and Sputnik V). This time window captures both the initial phase of the COVID-19 pandemic and the first vaccination roll-out. The case study offers a promising testing ground: both events were acknowledged during the COVID-19 pandemic as potential instances of information voids, despite no established quantitative metric existed to measure them~\cite{ishizumi2024beyond, lohiniva2022covid, briand2021infodemics, purnat2021infodemic}. Both demand and supply data were collected using keyword-based searches (see SI for details). In particular, Wikipedia page-views and Google Trends data were used as proxies for demand, while data from Facebook, Twitter, and GDELT were used as proxies for supply (each source was processed independently to maintain its distinct temporal and scale properties prior to rescaling).

\begin{figure}[ht!]
    \centering
    \includegraphics[width=1\linewidth]{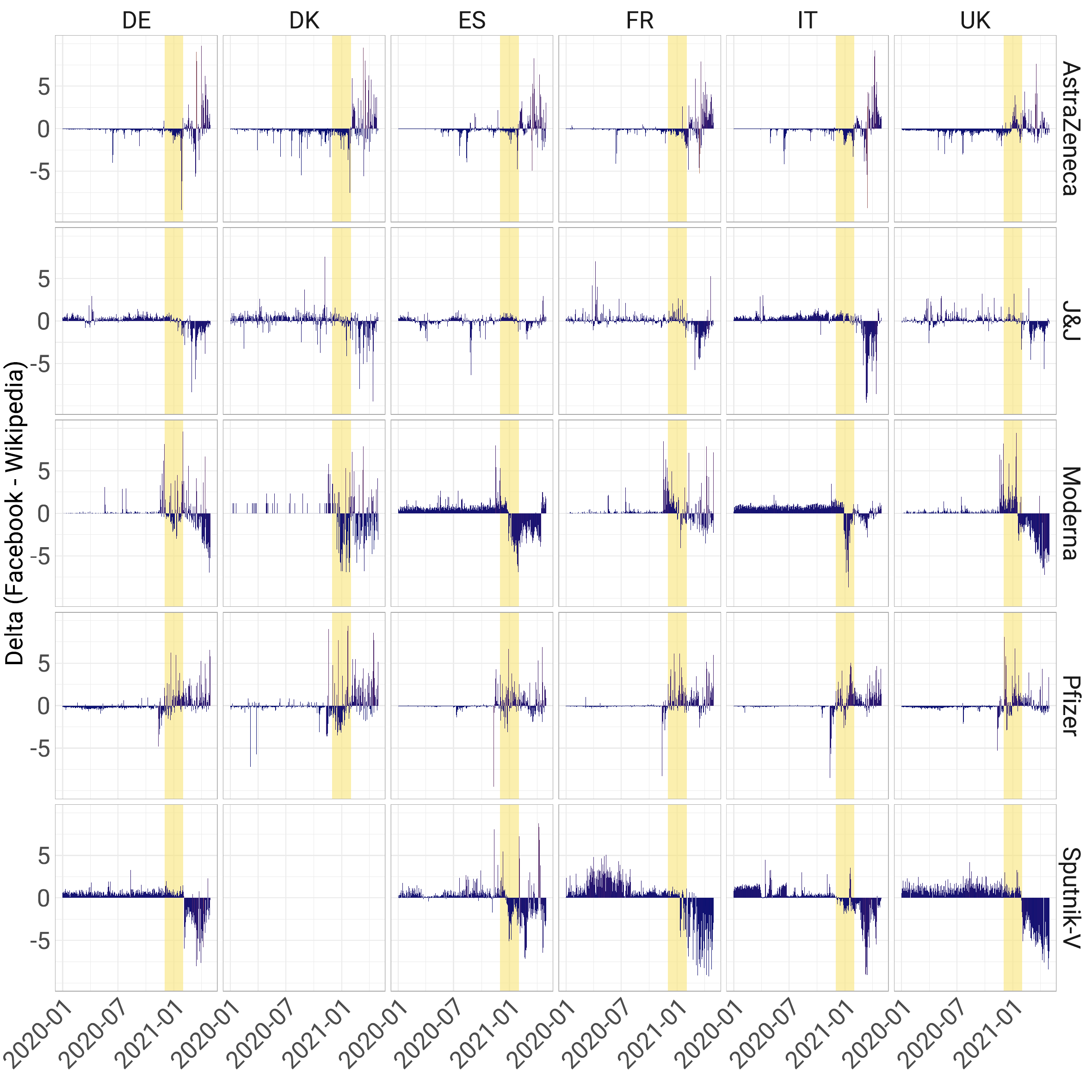}
    \caption{Daily variation of the supply-demand delta computing Facebook supply and Wikipedia demand. Each panel displays trends over time for a specific combination of country (column label) and vaccine (row label). Time series for $\delta_t$ were capped to $\pm$ 10 for visualisation purposes, yet much larger values, i.e. more severe anomalies, occur. The yellow-shaded region highlights the period from December 1, 2020, to February 1, 2021, encompassing the time immediately before and after the introduction of the vaccine. The empty panel corresponds to the Sputnik-V vaccine in Denmark, for which no information is available during the same time span.}
    \label{fig:delta_wiki_main}
\end{figure}

Figure~\ref{fig:delta_wiki_main} illustrates the evolution of the information delta for Facebook (supply) and Wikipedia (demand), while results for all the combination of platforms (including also Twitter and GDELT in the case of supply and Google Trends in the case of demand) are provided in SI and show consistent trends.  
Considering the beginning of the observation period, we note that $\delta_t$ remains relatively stable, suggesting a general balance between supply and demand. However, in the latter half, pronounced fluctuations appear, with both positive and negative peaks. Negative delta values reflect instances where information demand substantially exceeds supply, whereas a positive delta denotes the opposite pattern, with supply surpassing demand.
The most relevant fluctuations coincide temporally with the onset of vaccination campaigns in several countries and the  intensification of public debate surrounding the initial roll-out of COVID-19 vaccines. As hypothesized, external events closely related to the topics under investigation (vaccines) triggered measurable disruptions within the information landscape. These disruptions produced immediate shifts in both the volume of available content and the public’s information needs. The information delta, in this period, exhibits marked oscillations with a predominance of negative values, indicating difficulties in aligning content production with rising public demand during critical periods. Additional analyses based on Google Trends data, presented in the SI, corroborate these findings. 

\subsubsection*{Anomaly detection analysis}
\begin{figure}[ht!]
    \centering
    \includegraphics[width=1\linewidth]{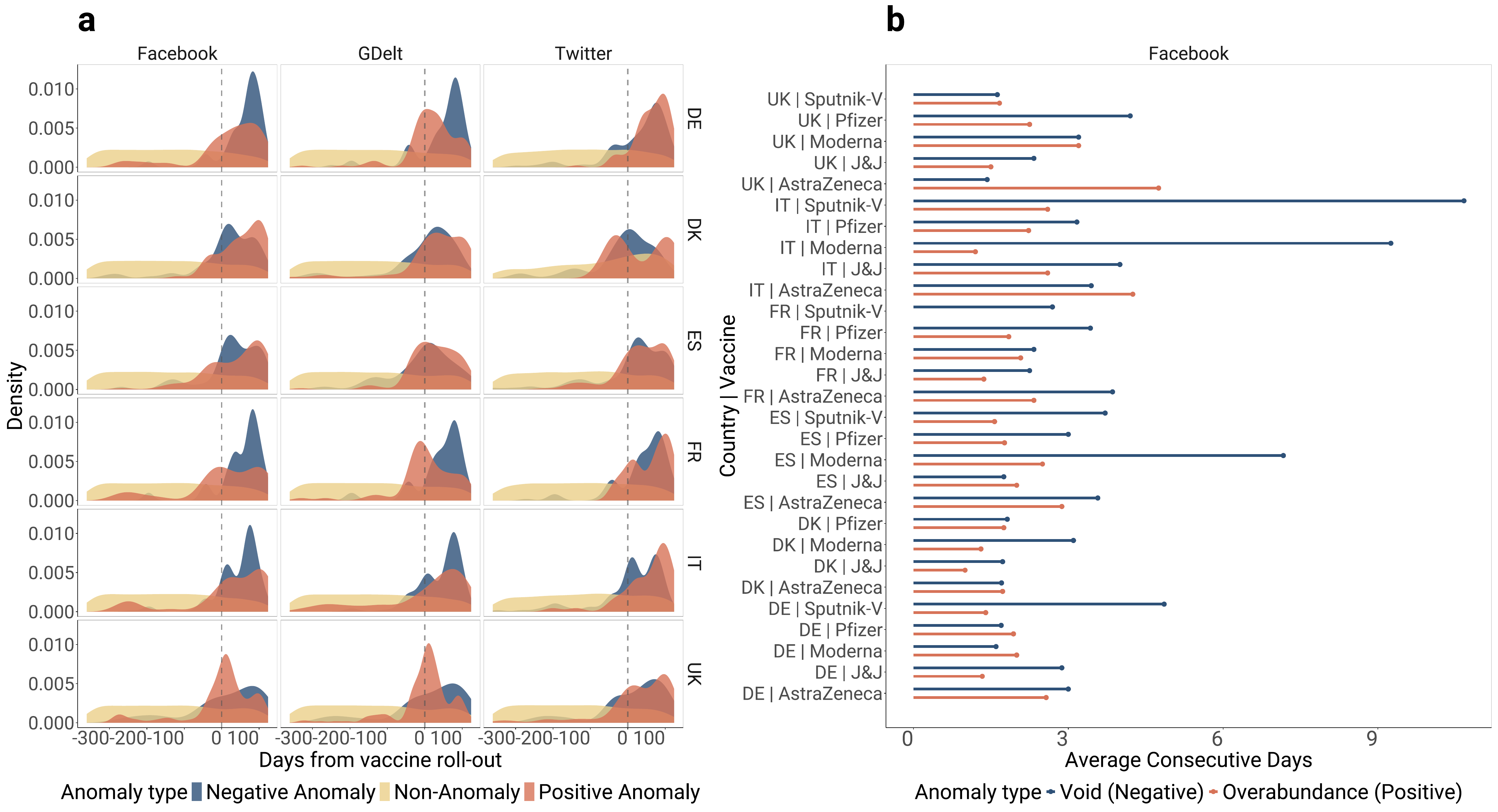}
    \caption{a) Distribution of anomalies before and after the vaccine roll-out. The x-axis represents days from the national start of the vaccination campaign (vertical dashed line at day 0). b) Average duration of anomalies by source. Negative anomalies are displayed in blue while positive anomalies in red.}
    \label{fig:density_anomaly}
\end{figure}

Although the information delta provides a reasonable approximation of supply–demand imbalances, our focus is on capturing extreme deviations in the delta's time series. We therefore rely on anomaly detection methods based on conservative statistical criteria to identify significant departures from typical dynamics, corresponding to information overabundance (positive) and information voids (negative). The results of this analysis are reported in Table~\ref{tab:anomaly_perc}, which summarizes the proportion of detected anomalies across platforms. The SI includes time series for all country–source–vaccine combinations with anomalies highlighted.
\begin{table}[ht!]
    \centering
    \begin{tabular}{l l c c c c c c c}
        Country & Source & Total & \% Anomalies & \% Positive & \% Negative \\ 
        \hline
        DE & Facebook & 2430 & 9.01 & 4.40 & 4.61 \\ 
        DE & GDelt & 2425 & 8.16 & 4.33 & 3.84 \\ 
        DE & Twitter & 2081 & 9.90 & 5.00 & 4.90 \\ 
        DK & Facebook & 1944 & 8.49 & 3.40 & 5.09 \\ 
        DK & GDelt & 1940 & 7.89 & 4.33 & 3.56 \\ 
        DK & Twitter & 909 & 9.79 & 2.09 & 7.70 \\ 
        ES & Facebook & 2430 & 8.64 & 3.46 & 5.19 \\ 
        ES & GDelt & 2425 & 7.88 & 4.54 & 3.34 \\ 
        ES & Twitter & 2292 & 9.90 & 5.41 & 4.49 \\ 
        FR & Facebook & 2430 & 8.35 & 3.58 & 4.77 \\ 
        FR & GDelt & 2425 & 8.78 & 4.74 & 4.04 \\ 
        FR & Twitter & 2148 & 9.87 & 4.93 & 4.93 \\ 
        UK & Facebook & 2430 & 7.98 & 4.94 & 3.05 \\ 
        UK & GDelt & 2425 & 8.37 & 6.14 & 2.23 \\ 
        UK & Twitter & 2422 & 9.87 & 6.69 & 3.18 \\ 
        IT & Facebook & 2430 & 9.42 & 3.21 & 6.21 \\ 
        IT & GDelt & 2425 & 7.92 & 3.92 & 4.00 \\ 
        IT & Twitter & 1971 & 9.84 & 4.11 & 5.73 \\ 
        \hline
    \end{tabular}
    \caption{Total number of observations of the information delta, along with the percentage of detected anomalies, and the breakdown of positive and negative anomalies, using Wikipedia page-views as a proxy for information demand.}
    \label{tab:anomaly_perc}
\end{table}
To quantify these deviations and provide an initial validation of the data science pipeline, we analyse the distribution of anomalies before and after the start of the vaccination campaign (Figure~\ref{fig:density_anomaly}a). 
The results indicate that, prior to the vaccine rollout, the information space remained largely stable, characterised by a predominance of non-anomalous behaviour. During this pre-rollout phase, both positive and negative anomalies were infrequent and weakly clustered, suggesting a general equilibrium between content production and information-seeking activity. This is expected and consistent with the observation that during the first phase of the pandemic, discussion on COVID-19 vaccines was largely generic and became more specific, including mentions of individual vaccine brands, as the vaccination roll-out approached. 
More specifically, for the supply proxies, 34.2\% of activity occurred in the pre-rollout period, while 65.8\% occurred post-rollout. For the demand proxies, 27\% of activity occurred pre-rollout and 73\% post-rollout (all country–source–vaccine combinations are reported in the SI). In contrast, around the onset of vaccination campaigns, all platforms exhibited a peak in the frequency of anomalous behaviour, marked by an overlapping concentration in both positive and negative anomalies. The timing and shape of these peaks display some degree of variation by platforms and countries: in many instances (e.g. GDELT in DE and Facebook and Twitter in UK) the peak of positive anomalies preceeds the negative ones while in other cases (e.g., Twitter in IT and and Facebook in ES) the opposite behaviour is observed. In general, and except for few exceptions like Facebook and GDELT in UK, negative anomalies (that is, information voids) display a more pronounced peak and thus a higher concentration and frequency than positive ones.
To further characterise the temporal dynamics of anomalies,  (Figure~\ref{fig:density_anomaly}b displays their average persistence measured in terms of consecutive "anomalous" days. In other words, persistence represents the duration of anomalous phases, indicating whether imbalances are transient or sustained, and it is defined as the number of consecutive, daily anomalies having the same sign. To allow for brief interruptions within otherwise continuous episodes, we applied a tolerance rule: if anomalies of the same sign were separated by at most two non-anomalous days, these days were included in the persistence calculation. 

In particular, using Wikipedia as a demand proxy, the longest periods of information void include 29 consecutive days (i.e., 29 consecutive anomalies). For instance, Twitter displays an anomaly lasting 29 days considering the case of Sputnik V vaccine in Germany, Facebook displays an anomaly lasting the same number of days for Moderna in Italy, while GDELT displays a 27-day anomaly for the Sputnik V vaccine in Germany. In contrast, the longest periods of overabundance were 26 consecutive days on Twitter in the United Kingdom for Johnson \& Johnson, 25 days on GDELT in France for Moderna, and 15 days on Facebook in the United Kingdom for Moderna. These results reveal a marked asymmetry: information voids tend to persist longer than periods of overabundance. This pattern suggests that the ecosystem returns to balance more slowly following sudden spikes in public demand than after episodes of excess content production. 
The SI reports the distribution of anomalies before and after the start of the vaccination campaign using Google Trends as a proxy for information demand, as well as the complete results on the average persistence of anomalies for both Wikipedia page-views and Google Trends as demand proxies.

\subsection*{Anomaly validation and correlation with misinformation} 
\subsubsection*{Case studies} 
As an additional investigation, we perform a focused analysis on two high-impact vaccine-related events that occurred during the study period. The inclusion of case studies allows us to examine in detail how debated events within the study period produce marked imbalances between information supply and demand. Specifically, we examine (a) the conditional authorization of the Moderna vaccine by the European Medicines Agency (EMA), and (b) the suspension and subsequent reinstatement of the AstraZeneca vaccine due to safety concerns. 

\begin{figure}[H]
    \centering
    \includegraphics[width=1\linewidth]{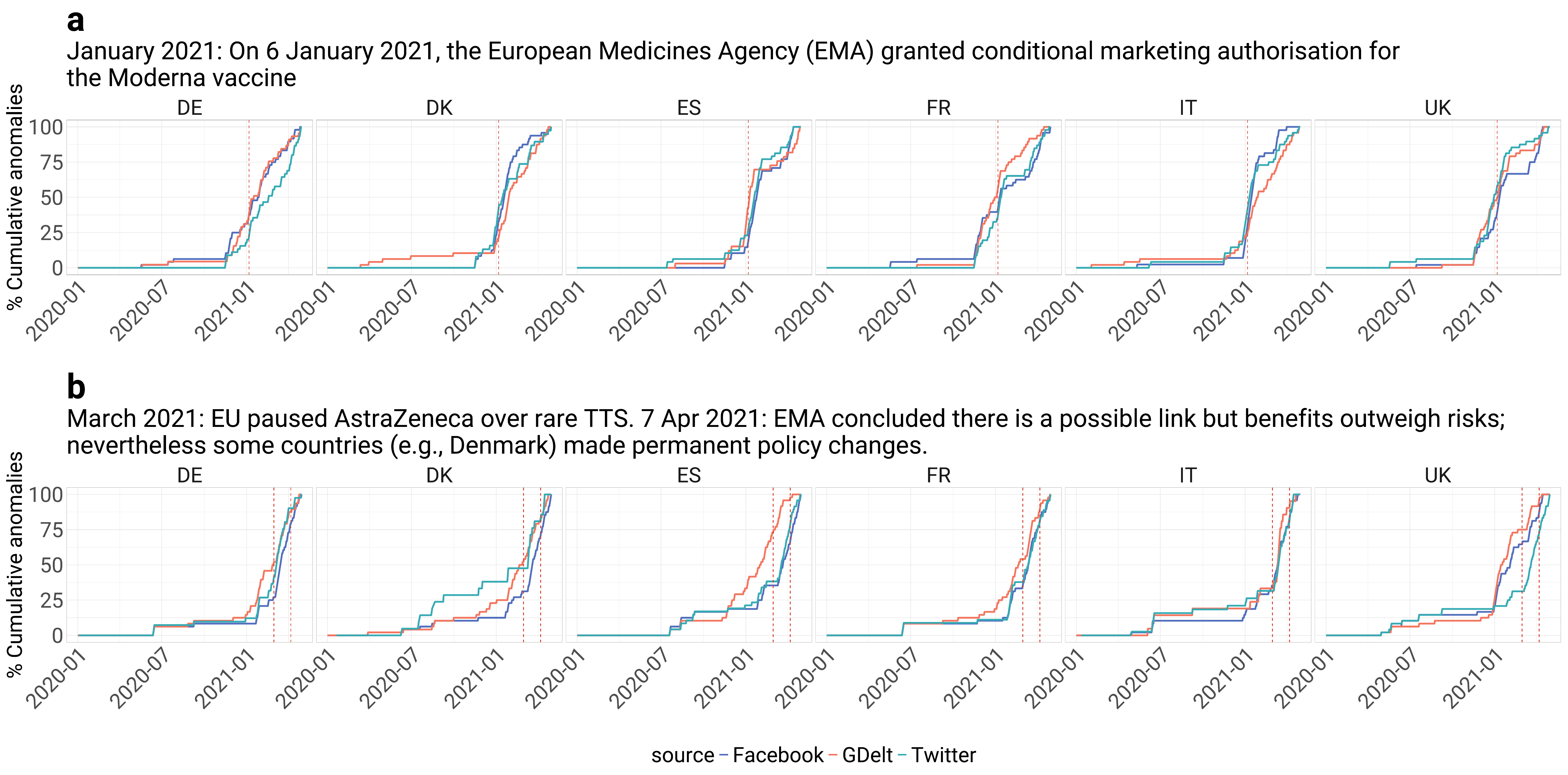}
    \caption{a) Cumulative value of anomalies for the Moderna vaccine (mrna-1273), displayed by source and country. The red line indicates the date of authorization by the EMA. b) Cumulative anomalies related to the AstraZeneca vaccine, broken down by source and country. The red vertical lines mark key dates: 1 March, when the first cases of TTS were being reported and the debate over AstraZeneca’s safety was beginning to intensify across several European states, and 7 April when the EMA Pharmacovigilance Risk Assessment Committee (PRAC) confirmed a causal link with TTS as a very rare adverse event.}
    \label{fig:CaseStudy}
\end{figure}

The first case study on the Moderna vaccine focuses on the period when the EMA authorized its distribution \cite{EMA_Moderna_2021}. Figure ~\ref{fig:CaseStudy}a illustrates that during this period, anomalies increased substantially, with this rise being persistent and widespread across all sources and countries analysed. In Germany, anomalies detected on Facebook accounted for 37\% of cumulative observations up to January 6 but one month after the vaccine approval - on February 6 - this share increased to 67\%. Similar patterns were observed elsewhere during the same period: Denmark increased from 33\% to 83\%, Spain from 18\% to 68\%, France from 41\% to 63\%, the United Kingdom from 39\% to 67\%, and Italy from 30\% to 81\%. This persistent rise indicates a widespread and consistent impact across all countries and sources examined.

The second case study (Figure~\ref{fig:CaseStudy}b) focuses on the period between March 1, 2021, when several EU member states initiated precautionary suspensions of the AstraZeneca vaccine following reports of thrombosis with thrombocytopenia syndrome (TTS) \cite{webster2021covid}, and April 7, 2021, when the EMA’s Pharmacovigilance Risk Assessment Committee (PRAC) confirmed a causal association between the AstraZeneca vaccine and TTS as a very rare adverse event \cite{WHO_GACVS_2021}. In particular, on 11 March, Denmark, Norway and Iceland suspended the use of the ChAdOx1 vaccine. During this period, a significant increase in anomalies was observed across all countries and sources. Again using Facebook as an illustrative example, cumulative anomalies in Germany increased from 27\% on 1 March to 79\% on 7 April. Denmark showed a similar trend, increasing from 31\% to 70\%, while Spain rose from 35\% to 67\%, France from 35\% to 82\%, the United Kingdom from 64\% to 87\%, and Italy from 35\% to 88\%. 

The SI reports the results of all case studies using Google Trends as a proxy for information demand which confirms what reported in the main paper. In addition, to the cases discussed in the main text, the SI also includes two further case studies related to the Pfizer vaccine (concerning reports by the Norwegian Medicines Agency on the investigation of deaths among elderly individuals following vaccination, which were widely misinterpreted as vaccine-related) and the Johnson \& Johnson vaccine, following the recommendation by the CDC and FDA to pause its use due to rare blood clotting events.
\subsubsection*{Relationship between anomalies and reduction of information quality} 
We investigate whether the presence of anomalies correlates with a deterioration in the quality of information circulating in the system. In particular, anomalies in the information delta time series, and especially information voids, may mark periods of increased prevalence of misinformation \cite{wehrli2025information, nature2024_infovoids}.

To explore this aspect, we combine Facebook and Twitter dataset (and Wikipedia as demand proxy) with data from NewsGuard in order to obtain credibility scores for online sources and we analyse how average credibility scores varies across periods of non-anomaly, void (negative anomaly) and overabundance (positive anomaly). In particular, NewsGuard evaluates news and information websites based on nine apolitical criteria that assess their credibility and transparency. The overall score ranges from 0 to 100, where sites scoring 75 or higher are classified as generally credible, those between 60 and 74 as credible with exceptions, between 40 and 59 as "proceed with caution", and below 40 as "proceed with maximum caution". After matching NewsGuard credibility scores of websites with URL domains posted on Facebook and Twitter, the resulting datasets included 320,539 Facebook posts and 2,254,309 Twitter posts containing a labelled domain. Note that the reduction in the data compared to the initial dataset, is primarily due to the exclusion of posts not containing any URL or containing a domain not rated by NewsGuard.
As an initial observation, the average news reliability on Twitter (\(\approx\)70) is lower than that on Facebook (\(\approx\)84). After this initial investigation, we analysed the distribution of posts according to NewsGuard’s reliability classification to assess how information quality varies in correspondence of different anomaly conditions. Table ~\ref{tab:tab_anomaly_score} reports percentages of labelled domains by social media platform, calculated from December 8, 2020, corresponding to the launch of the COVID-19 vaccination campaign in the United Kingdom and the onset of increased anomalous activity. 

In particular, during periods of information voids (i.e., negative anomalies), content from highly credible sources (S $=100$) accounts for only 20.84\% of posts on Facebook and 23.87\% of posts on Twitter, considerably lower than the 33.68\% and 31.62\% observed on Facebook and Twitter, respectively, during non-anomalous periods. These percentages are also lower than those observed during positive anomalies (28.59\% on Facebook and 31.07\% on Twitter), suggesting that during periods of exceptionally high demand of information, users have reduced access to high-quality content within the information ecosystem. We also observe an increase in the presence of content labelled as “proceed with maximum caution” (S $< 39$, i.e. misinformation) during periods of information voids. Specifically, such content accounts for 6.1\% of Facebook posts during voids (compared to 5.46\% in non-anomalous periods and 4.5\% during positive anomalies) and for 27.28\% of Twitter content (compared to 22.1\% in non-anomalous periods and 15.69\% during positive anomalies). Taken together with the simultaneous decline in highly reliable information, these patterns highlight information voids as periods in which access to reliable content becomes more limited, while prevalence of highly unreliable information increases.
\begin{table}[ht!]
    \centering
    \begin{tabular}{l|c c c c c|c}
        \textbf{Anomaly type} & \textbf{S = 100} & \textbf{99$<$S$<$75} & \textbf{74$<$S$<$60} & \textbf{59$<$S$<$40} & \textbf{S$<$39} & \textbf{Source} \\
        \hline
        Non-anomaly        & 33.68\% & 51.37\% & 7.5\% & 1.8\% & 5.46\% & Facebook \\
        Negative anomaly   & 20.84\% & 59.17\% & 11.47\% & 2.05\% & 6.1\% & Facebook \\
        Positive anomaly   & 28.59\% & 56.28\% & 9.12\% & 1.32\% & 4.5\% & Facebook \\
        \hline
        Non-anomaly        & 31.62\% & 37.18\% & 6.87\% & 1.28\% & 22.51\% & Twitter \\
        Negative anomaly   & 23.87\% & 40.22\% & 6.85\% & 1.23\% & 27.28\% & Twitter \\
        Positive anomaly   & 31.07\% & 43.66\% & 7.25\% & 1.72\% & 15.69\% & Twitter \\
    \end{tabular}
    \caption{Percentage of posts published by score based on NewsGuard classification.}
    \label{tab:tab_anomaly_score}
\end{table}

\section*{Conclusions}\label{sec4}
The online information ecosystem plays a crucial role today as it represents one of the primary sources through which users access news \cite{newman2025digital} and develop opinions. 
Within this context, misinformation plays a crucial role, yet many of its drivers have not been fully understood \cite{van2022misinformation, vraga2020defining}. 
Beyond providing a method for detecting information voids, our study shows that such regimes of information production and demand may play a role in determining the prevalence of misinformation and the reduction of the overall quality of information circulating online. Indeed, both the proportion of content produced by highly unreliable and reliable sources vary in opposite directions during information voids. 

In the scientific literature, the production and consumption of misinformation have been predominantly studied from a user-centered perspective \cite{pennycook2021psychology,budak2024misunderstanding}, without offering a macroscopic explanation for the emergence of such phenomena.
While our study does not establish a causal link between the presence of information voids and increases in misinformation, this association is clearly observable in the analysed data and could be considered into models aimed at providing a mechanistic explanation of misinformation emergence. More specifically, the onset of the COVID-19 vaccine debate and the subsequent rollout of vaccination campaigns constituted significant perturbations to the information ecosystem and were associated with pronounced peaks in information anomalies.

Another contribution stemming from this study is the development of a classification of the regimes of information supply and demand, which allows for the systematic and replicable identification of different phases of equilibrium or imbalance in the system. The classification includes five regimes: \textit{Void}; \textit{Lack}; \textit{Balance}; \textit{Abundance}; \textit{Overabundance}. This classification offers a valuable tool for monitoring and potentially anticipating risk conditions for misinformation spread, with important implications for both research and policy interventions.

Beyond the methodological focus of this work, our framework has direct implications for communication preparedness in public health. By enabling the systematic, quantitative characterisation of supply/demand imbalances, it provides an early indication of stress within the information ecosystem, complementing taxonomy-based, expert-driven social listening approaches that detect infodemic signals and potential information voids through the analysis of circulating content \cite{purnat2021infodemic}. As a quantitative layer that can be integrated into existing risk communication and infodemic management frameworks \cite{ECDC2025_preparedness,world2022advancing, veil2008cerc}, the model can inform the prioritization and timing of institutional communication efforts during critical phases of a health crisis. In addition, tracking information regimes over time, together with their association with the circulation of misinformation, provides a basis for ex post evaluation of whether communication responses effectively mitigated periods of informational vulnerability.

Regarding limitations, it is important to emphasize that, despite our approach leverages multiple platforms, our dataset is not fully representative of the general public in terms of demographics, interests, and behaviours. Information on vaccinations and, more generally, on the COVID-19 pandemic has also been disseminated through a wide range of other channels, including traditional media and alternative online platforms. Nonetheless, the analysis of news through GDELT reveals an information supply deficit, confirming that this imbalance is not limited to social media platforms alone. It should be stressed that, despite we used two proxies for information demand, the online information ecosystem is not the only avenue through which individuals access news and form opinions - either during the pandemic or in everyday life more broadly. Nevertheless, the primary aim of this work was to provide a robust and reproducible data science pipeline for defining and measuring the overall health of the online information ecosystem and assessing its ability to respond to users’ informational demands. In doing so, we were able to identify when the feedback loop between supply and demand risks breaking down, thereby exposing areas that are more vulnerable to misinformation.

\section*{Methods}\label{sec:Methods}
Data on both supply and demand were collected through a keyword-based search strategy using a consistent set of terms across all platforms and tools. The keywords, refers to the five main COVID-19 vaccines: AstraZeneca, Moderna, Johnson \& Johnson, Pfizer, and Sputnik V. 

\subsection*{Supply side data}\label{sec3.1}
Data from Facebook and Twitter were collected for the period between 1 January 2020 and 30 April 2021. Facebook posts were obtained through CrowdTangle \cite{crowdtangle2024}, a Meta-owned data analysis platform that remained active until 14 August 2024. A keyword-based search strategy was implemented to identify content related to vaccines and associated brands. Following the same approach adopted for Facebook, we collected data from Twitter by means of a full-archive historic search within the v2 endpoint and academic research product track. Data on online news articles were collected using the GDELT Summary platform, made available as part of the GDELT project. This advanced research tool enables exploration of textual and visual narratives from global news media in multiple languages. Within the platform, customized dashboards can be generated to synthesize global coverage related to specific topics. For this study, a keyword-based search was performed on the Global Online News Coverage stream, filtering results by the countries of interest. The keywords used for GDELT extraction were aligned with those employed in the Facebook and Twitter data collection.

The final dataset comprises 1.040.624 Facebook posts, 35.164.141 Twitter posts, and 1.343.052 online news articles from GDELT. The complete list of keywords used for all three data sources is provided in the SI.

\subsection*{Demand side data}\label{sec3.2}
To examine the demand for information, two distinct sources were used. Data were collected for all countries under study, covering the period from January 1, 2020, to April 30, 2021.

The first proxy for information demand is derived from Wikipedia page views. Data were retrieved using the Pageviews Analysis tool \cite{wikimedia_pageviews}, applying a keyword search strategy consistent with that used to measure information supply. Wikipedia provides page-level data at daily resolution, yielding time series that record the number of visits to each page on a day-by-day basis. It is important to note that for Denmark, page-view data related to Sputnik V are not available.

The second proxy for information demand is Google Trends. We used the \texttt{gtrendsR} package in R \cite{massicotte2016package}, which enables keyword-based searches with strings of terms aligned with those commonly adopted on social media. This choice ensured control over the terminology embedded in the broader reference topics, particularly vaccines. To ensure consistency, the keyword selection for Google Trends was harmonized with that applied across all other data sources. For each topic, Google Trends returns a weekly time series normalized on a 0–100 scale, where 100 corresponds to the peak search interest within the selected period. A value of 0 does not necessarily indicate the absence of searches, but rather a volume too low to be reported. 
The complete list of keywords used for Wikipedia Page-views and Google Trends is provided in the SI.

The temporal granularity of data differs across data sources: Google Trends provides data at weekly resolution, whereas Wikipedia and the supply-side platforms provide daily data. Consequently, different procedures were applied to harmonize the datasets. Specifically, when comparing supply data with demand data obtained from Google Trends, weekly social media post volumes were aggregated and rescaled so that the maximum weekly value corresponds to 100, mirroring the Google Trends scale. The rescaled weekly supply $\hat{s}_w$ is defined as:
\begin{equation}
\underline{s}_w = \biggl\lfloor \frac{s_w}{\max_{w \in W} s_w} \cdot 100\biggr\rfloor
\label{eq:norm_supply}
\end{equation}
where $s_w$ is the number of posts in week $w$, and $W$ is the set of all weeks in the observation period. 

\subsection*{Classification of media sources}
To assign reliability labels to posts published on online platforms, we relied on the classification of web pages provided by NewsGuard, an independent agency that evaluates the credibility of news sources. NewsGuard employs a team of experienced journalists and editors to assess websites that publish news and information, assigning ratings and reliability scores based on nine journalistic criteria \cite{NewsGuard}. These criteria reflect fundamental standards of credibility and transparency in journalism. Each criterion contributes to a cumulative score ranging from 0 to 100, reported as a percentage, which represents the overall reliability of the site. Higher scores indicate stronger adherence to journalistic standards, with scores above 60 generally denoting trustworthy sources. 

\subsection*{Anomaly detection} \label{Anomaly Detection}
Algorithm~\ref{alg:anomaly_det} outlines the unsupervised anomaly detection procedure employed in this study \cite{dancho2020package}. The approach comprises three main steps: time-series decomposition, anomaly detection on the remainder component (via the IQR method), and subsequent time-series recomposition. 
Time series decomposition techniques aim to separate a time series into multiple components, including trend (long-term movement), seasonality (periodic fluctuations, weekly in the case of daily data) and a remainder component capturing residual variation. In this study, we employ the STL (Seasonal–Trend decomposition using Loess) method \cite{cleveland1990stl}, which performs an additive decomposition of the series through iterative applications of the Loess smoother \cite{theodosiou2011forecasting}.
Additional methodological details are provided in the SI.

\begin{algorithm}[H]
\caption{Anomaly Detection on Information Delta} 
\label{alg:anomaly_det}
\begin{algorithmic}[1]
\State \textbf{Input:} Time series of information delta $\{\delta_t\}_{t=1}^T$
\State \textbf{Output:} Anomaly labels $\mathcal{A}_t \in \{\text{Yes}, \text{No}\}$
\State \textbf{Step 1: Time series decomposition}: decompose $\delta_t$ into seasonal, trend, and remainder components:
\[
\delta_t = S_t + T_t + R_t
\]
where $S_t$ represents the seasonal (cyclic) component, $T_t$ the long-term trend, and $R_t$ the remainder.
\State Compute the remainder explicitly:
\[
R_t = \delta_t - (S_t + T_t)
\]

\State \textbf{Step 2: Anomaly detection on remainder}: Define $R$ as the empirical distribution of $R_t$ values with $t \in [1,T]$ and compute the interquartile range (IQR) of $R$:
\[
IQR = Q_3(R) - Q_1(R)
\]
$Q_1$ and $Q_3$ are the first and the third quartiles of $R$.

\State Define anomaly thresholds:
\[
L_1 = Q_1(R) - k(\alpha) \cdot IQR, \quad
L_2 = Q_3(R) + k(\alpha) \cdot IQR
\]
where $k(\alpha)$ is a parametric function used to define anomaly thresholds.
\State Label anomalies based on remainder:
\[
\mathcal{A}_t =
\begin{cases}
\text{Yes}, & \text{if } R_t < L_1 \text{ or } R_t > L_2 \\
\text{No}, & \text{otherwise}
\end{cases}
\]

\State \textbf{Step 3: Time Recomposition}: recompose the anomaly thresholds into the observed scale:
\[
\text{Recomposed}_t^{(L1)} = S_t + T_t + L_1, \quad
\text{Recomposed}_t^{(L2)} = S_t + T_t + L_2
\]
\State Observations outside these bounds are flagged as anomalous in the original series.

\State \textbf{Return} Anomaly labels $\mathcal{A}_t$ and recomposed limits.
\end{algorithmic}
\end{algorithm}



\bigskip

\bibliography{sn-bibliography}

@article{gravino2024online,
  title={Online news ecosystem dynamics: supply, demand, diffusion, and the role of disinformation},
  author={Gravino, Pietro and Prevedello, Giulio and Brugnoli, Emanuele},
  journal={Applied Network Science},
  volume={9},
  number={1},
  pages={40},
  year={2024},
  publisher={Springer}
}

@article{gravino2022supply,
  title={The supply and demand of news during COVID-19 and assessment of questionable sources production},
  author={Gravino, Pietro and Prevedello, Giulio and Galletti, Martina and Loreto, Vittorio},
  journal={Nature human behaviour},
  volume={6},
  number={8},
  pages={1069--1078},
  year={2022},
  publisher={Nature Publishing Group UK London}
}

@article{scalco2025modelling,
  title={Modelling the Climate Change Debate in Italy through Information Supply and Demand},
  author={Scalco, Irene and Colafrancesco, Giulia and Cinelli, Matteo},
  journal={arXiv preprint arXiv:2503.17026},
  year={2025}
}

@inproceedings{yoshida2015wikipedia,
  title={Wikipedia page view reflects web search trend},
  author={Yoshida, Mitsuo and Arase, Yuki and Tsunoda, Takaaki and Yamamoto, Mikio},
  booktitle={Proceedings of the ACM web science conference},
  pages={1--2},
  year={2015}
}

@book{hawkins1980identification,
  title={Identification of outliers},
  author={Hawkins, Douglas M},
  volume={11},
  year={1980},
  publisher={Springer}
}

@article{blazquez2021review,
  title={A review on outlier/anomaly detection in time series data},
  author={Bl{\'a}zquez-Garc{\'\i}a, Ane and Conde, Angel and Mori, Usue and Lozano, Jose A},
  journal={ACM computing surveys (CSUR)},
  volume={54},
  number={3},
  pages={1--33},
  year={2021},
  publisher={ACM New York, NY, USA}
}

@article{purnat2021digital,
  title={WHO Digital Intelligence Analysis for Tracking Narratives and Information Voids in the COVID-19 Infodemic},
  author={Purnat, Tina D and Vacca, Paolo and Burzo, Stefano and Zecchin, Tim and Wright, Amy and Briand, Sylvie and Nguyen, Tim},
  journal={Studies in Health Technology and Informatics},
  volume={281},
  pages={989--993},
  year={2021},
  doi={10.3233/SHTI210326}
}

@article{nature2024_infovoids,
  title        = {How online misinformation exploits ‘information voids’ — and what to do about it},
  author       = {{Nature Editorial}},
  journal      = {Nature},
  volume       = {625},
  pages        = {215--216},
  year         = {2024},
  month        = jan,
  day          = {9},
  doi          = {10.1038/d41586-024-00030-x},
  url          = {https://doi.org/10.1038/d41586-024-00030-x},
}

@article{purnat2021infodemic,
  title={Infodemic signal detection during the COVID-19 pandemic: development of a methodology for identifying potential information voids in online conversations},
  author={Purnat, Tina D and Vacca, Paolo and Czerniak, Christine and Ball, Sarah and Burzo, Stefano and Zecchin, Tim and Wright, Amy and Bezbaruah, Supriya and Tanggol, Faizza and Dub{\'e}, {\`E}ve and others},
  journal={JMIR infodemiology},
  volume={1},
  number={1},
  pages={e30971},
  year={2021},
  publisher={JMIR Publications Toronto, Canada}
}

@article{cebrian2023google,
  title={Is Google Trends a quality data source?},
  author={Cebri{\'a}n, Eduardo and Domenech, Josep},
  journal={Applied Economics Letters},
  volume={30},
  number={6},
  pages={811--815},
  year={2023},
  publisher={Taylor \& Francis}
}

@article{costola2021google,
  title={Google search volumes and the financial markets during the COVID-19 outbreak},
  author={Costola, Michele and Iacopini, Matteo and Santagiustina, Carlo RMA},
  journal={Finance Research Letters},
  volume={42},
  pages={101884},
  year={2021},
  publisher={Elsevier}
}

@article{jun2018ten,
  title={Ten years of research change using Google Trends: From the perspective of big data utilizations and applications},
  author={Jun, Seung-Pyo and Yoo, Hyoung Sun and Choi, San},
  journal={Technological forecasting and social change},
  volume={130},
  pages={69--87},
  year={2018},
  publisher={Elsevier}
}

@article{zhang2018using,
  title={Using Google Trends and ambient temperature to predict seasonal influenza outbreaks},
  author={Zhang, Yuzhou and Bambrick, Hilary and Mengersen, Kerrie and Tong, Shilu and Hu, Wenbiao},
  journal={Environment international},
  volume={117},
  pages={284--291},
  year={2018},
  publisher={Elsevier}
}

@techreport{dataandsociety2019datavoids,
  author = {Michele Golebiewski and danah boyd},
  title = {Data Voids: Where Missing Data Can Easily Be Exploited},
  institution = {Data \& Society},
  year = {2019},
  url = {https://datasociety.net/library/data-voids/}
}

@article{aslett2024online,
  title={Online searches to evaluate misinformation can increase its perceived veracity},
  author={Aslett, Kevin and Sanderson, Zeve and Godel, William and Persily, Nathaniel and Nagler, Jonathan and Tucker, Joshua A},
  journal={Nature},
  volume={625},
  number={7995},
  pages={548--556},
  year={2024},
  publisher={Nature Publishing Group UK London}
}

@article{avalle2024persistent,
  title={Persistent interaction patterns across social media platforms and over time},
  author={Avalle, Michele and Di Marco, Niccol{\`o} and Etta, Gabriele and Sangiorgio, Emanuele and Alipour, Shayan and Bonetti, Anita and Alvisi, Lorenzo and Scala, Antonio and Baronchelli, Andrea and Cinelli, Matteo and others},
  journal={Nature},
  volume={628},
  number={8008},
  pages={582--589},
  year={2024},
  publisher={Nature Publishing Group UK London}
}

@article{cinelli2021echo,
  title={The echo chamber effect on social media},
  author={Cinelli, Matteo and De Francisci Morales, Gianmarco and Galeazzi, Alessandro and Quattrociocchi, Walter and Starnini, Michele},
  journal={Proceedings of the National Academy of Sciences},
  volume={118},
  number={9},
  pages={e2023301118},
  year={2021},
  publisher={National Academy of Sciences}
}

@article{lazer2018science,
  title={The science of fake news},
  author={Lazer, David MJ and Baum, Matthew A and Benkler, Yochai and Berinsky, Adam J and Greenhill, Kelly M and Menczer, Filippo and Metzger, Miriam J and Nyhan, Brendan and Pennycook, Gordon and Rothschild, David and others},
  journal={Science},
  volume={359},
  number={6380},
  pages={1094--1096},
  year={2018},
  publisher={American Association for the Advancement of Science}
}

@article{tucker2018social,
  title={Social media, political polarization, and political disinformation: A review of the scientific literature},
  author={Tucker, Joshua A and Guess, Andrew and Barber{\'a}, Pablo and Vaccari, Cristian and Siegel, Alexandra and Sanovich, Sergey and Stukal, Denis and Nyhan, Brendan},
  journal={Political polarization, and political disinformation: a review of the scientific literature (March 19, 2018)},
  year={2018}
}

@article{del2016spreading,
  title={The spreading of misinformation online},
  author={Del Vicario, Michela and Bessi, Alessandro and Zollo, Fabiana and Petroni, Fabio and Scala, Antonio and Caldarelli, Guido and Stanley, H Eugene and Quattrociocchi, Walter},
  journal={Proceedings of the National Academy of Sciences},
  volume={113},
  number={3},
  pages={554--559},
  year={2016},
  publisher={National Academy of Sciences}
}

@article{allcott2019trends,
  title={Trends in the diffusion of misinformation on social media},
  author={Allcott, Hunt and Gentzkow, Matthew and Yu, Chuan},
  journal={Research \& politics},
  volume={6},
  number={2},
  pages={2053168019848554},
  year={2019},
  publisher={SAGE Publications Sage UK: London, England}
}

@article{van2017inoculating,
  title={Inoculating the public against misinformation about climate change},
  author={Van der Linden, Sander and Leiserowitz, Anthony and Rosenthal, Seth and Maibach, Edward},
  journal={Global challenges},
  volume={1},
  number={2},
  pages={1600008},
  year={2017},
  publisher={Wiley Online Library}
}

@article{alibudbud2023wikipedia,
  title={Wikipedia page views for health research: a review},
  author={Alibudbud, Rowalt},
  journal={Frontiers in Big Data},
  volume={6},
  pages={1199060},
  year={2023},
  publisher={Frontiers Media SA}
}

@article{moat2013quantifying,
  title={Quantifying Wikipedia usage patterns before stock market moves},
  author={Moat, Helen Susannah and Curme, Chester and Avakian, Adam and Kenett, Dror Y and Stanley, H Eugene and Preis, Tobias},
  journal={Scientific reports},
  volume={3},
  number={1},
  pages={1801},
  year={2013},
  publisher={Nature Publishing Group UK London}
}

@article{briand2021infodemics,
  title={Infodemics: A new challenge for public health},
  author={Briand, Sylvie C and Cinelli, Matteo and Nguyen, Tim and Lewis, Rosamund and Prybylski, Dimitri and Valensise, Carlo M and Colizza, Vittoria and Tozzi, Alberto Eugenio and Perra, Nicola and Baronchelli, Andrea and others},
  journal={Cell},
  volume={184},
  number={25},
  pages={6010--6014},
  year={2021},
  publisher={Elsevier}
}

@article{tangcharoensathien2020framework,
  title={Framework for managing the COVID-19 infodemic: methods and results of an online, crowdsourced WHO technical consultation},
  author={Tangcharoensathien, Viroj and Calleja, Neville and Nguyen, Tim and Purnat, Tina and D’Agostino, Marcelo and Garcia-Saiso, Sebastian and Landry, Mark and Rashidian, Arash and Hamilton, Clayton and AbdAllah, Abdelhalim and others},
  journal={Journal of medical Internet research},
  volume={22},
  number={6},
  pages={e19659},
  year={2020},
  publisher={JMIR Publications Inc., Toronto, Canada}
}

@article{cinelli2020covid,
  title={The COVID-19 social media infodemic},
  author={Cinelli, Matteo and Quattrociocchi, Walter and Galeazzi, Alessandro and Valensise, Carlo Michele and Brugnoli, Emanuele and Schmidt, Ana Lucia and Zola, Paola and Zollo, Fabiana and Scala, Antonio},
  journal={Scientific reports},
  volume={10},
  number={1},
  pages={16598},
  year={2020},
  publisher={Nature Publishing Group UK London}
}

@article{aldayel2021stance,
  title={Stance detection on social media: State of the art and trends},
  author={AlDayel, Abeer and Magdy, Walid},
  journal={Information Processing \& Management},
  volume={58},
  number={4},
  pages={102597},
  year={2021},
  publisher={Elsevier}
}

@article{loomba2021measuring,
  title={Measuring the impact of COVID-19 vaccine misinformation on vaccination intent in the UK and USA},
  author={Loomba, Sahil and De Figueiredo, Alexandre and Piatek, Simon J and De Graaf, Kristen and Larson, Heidi J},
  journal={Nature human behaviour},
  volume={5},
  number={3},
  pages={337--348},
  year={2021},
  publisher={Nature Publishing Group UK London}
}

@article{gallotti2020assessing,
  title={Assessing the risks of ‘infodemics’ in response to COVID-19 epidemics},
  author={Gallotti, Riccardo and Valle, Francesco and Castaldo, Nicola and Sacco, Pierluigi and De Domenico, Manlio},
  journal={Nature human behaviour},
  volume={4},
  number={12},
  pages={1285--1293},
  year={2020},
  publisher={Nature Publishing Group UK London}
}

@misc{wikimedia_pageviews,
  author       = {Wikimedia Foundation},
  title        = {Pageviews Analysis},
  howpublished = {\url{https://pageviews.wmcloud.org/}}
}

@misc{gdelt_summary_api,
  author       = {{The GDELT Project}},
  title        = {GDELT Summary API v2},
  howpublished = {\url{https://api.gdeltproject.org/api/v2/summary/}}
}

@article{nghiem2016analysis,
  title={Analysis of the capacity of Google Trends to measure interest in conservation topics and the role of online news},
  author={Nghiem, Le TP and Papworth, Sarah K and Lim, Felix KS and Carrasco, Luis R},
  journal={PloS one},
  volume={11},
  number={3},
  pages={e0152802},
  year={2016},
  publisher={Public Library of Science San Francisco, CA USA}
}

@article{liu2023anomaly,
  title={Anomaly and change point detection for time series with concept drift},
  author={Liu, Jiayi and Yang, Donghua and Zhang, Kaiqi and Gao, Hong and Li, Jianzhong},
  journal={World Wide Web},
  volume={26},
  number={5},
  pages={3229--3252},
  year={2023},
  publisher={Springer}
}

@article{schmidl2022anomaly,
  title={Anomaly detection in time series: a comprehensive evaluation},
  author={Schmidl, Sebastian and Wenig, Phillip and Papenbrock, Thorsten},
  journal={Proceedings of the VLDB Endowment},
  volume={15},
  number={9},
  pages={1779--1797},
  year={2022},
  publisher={VLDB Endowment}
}

@inproceedings{teng2010anomaly,
  title={Anomaly detection on time series},
  author={Teng, Mingyan},
  booktitle={2010 IEEE International Conference on Progress in Informatics and Computing},
  volume={1},
  pages={603--608},
  year={2010},
  organization={IEEE}
}

@article{van2022misinformation,
  title={Misinformation: susceptibility, spread, and interventions to immunize the public},
  author={Van Der Linden, Sander},
  journal={Nature medicine},
  volume={28},
  number={3},
  pages={460--467},
  year={2022},
  publisher={Nature Publishing Group}
}

@article{bovet2019influence,
  title={Influence of fake news in Twitter during the 2016 US presidential election},
  author={Bovet, Alexandre and Makse, Hern{\'a}n A},
  journal={Nature communications},
  volume={10},
  number={1},
  pages={7},
  year={2019},
  publisher={Nature Publishing Group UK London}
}

@article{vosoughi2018spread,
  title={The spread of true and false news online},
  author={Vosoughi, Soroush and Roy, Deb and Aral, Sinan},
  journal={Science},
  volume={359},
  number={6380},
  pages={1146--1151},
  year={2018},
  publisher={American Association for the Advancement of Science},
  url={https://science.sciencemag.org/content/359/6380/1146}
}

@article{schmidt2017anatomy,
  title={Anatomy of news consumption on Facebook},
  author={Schmidt, Ana Luc{\'\i}a and Zollo, Fabiana and Del Vicario, Michela and Bessi, Alessandro and Scala, Antonio and Caldarelli, Guido and Stanley, H Eugene and Quattrociocchi, Walter},
  journal={Proceedings of the National Academy of Sciences},
  volume={114},
  number={12},
  pages={3035--3039},
  year={2017},
  publisher={National Academy of Sciences}
}

@article{chou2018addressing,
  title={Addressing health-related misinformation on social media},
  author={Chou, Wen-Ying Sylvia and Oh, April and Klein, William MP},
  journal={Jama},
  volume={320},
  number={23},
  pages={2417--2418},
  year={2018},
  publisher={American Medical Association}
}

@article{bak2022combining,
  title={Combining interventions to reduce the spread of viral misinformation},
  author={Bak-Coleman, Joseph B and Kennedy, Ian and Wack, Morgan and Beers, Andrew and Schafer, Joseph S and Spiro, Emma S and Starbird, Kate and West, Jevin D},
  journal={Nature Human Behaviour},
  volume={6},
  number={10},
  pages={1372--1380},
  year={2022},
  publisher={Nature Publishing Group UK London}
}

@article{vraga2020defining,
  title={Defining misinformation and understanding its bounded nature: Using expertise and evidence for describing misinformation},
  author={Vraga, Emily K and Bode, Leticia},
  journal={Political Communication},
  volume={37},
  number={1},
  pages={136--144},
  year={2020},
  publisher={Taylor \& Francis}
}

@book{newman2025digital,
  title={Digital news report 2025},
  author={Newman, Nic and Ross Arguedas, A and Robertson, Craig T and Nielsen, Rasmus Kleis and Fletcher, Richard},
  year={2025},
  publisher={Reuters Institute for the study of Journalism}
}

@article{wehrli2025information,
  title={Information Pathways and Voids in Critical German Online Communities During the COVID-19 Vaccination Discourse: Cross-Platform and Mixed Methods Analysis},
  author={Wehrli, Silvan and Hartner, Anna-Maria and Boender, T Sonia and Arnrich, Bert and Irrgang, Christopher},
  journal={Journal of Medical Internet Research},
  volume={27},
  pages={e76309},
  year={2025},
  publisher={JMIR Publications Toronto, Canada}
}

@misc{NewsGuard,
  author       = {NewsGuard Technologies, Inc.},
  title        = {Website Rating Process and Criteria},
  year         = {2023},
  url          = {https://www.newsguardtech.com/ratings/rating-process-criteria/}
}

@article{lohiniva2022covid,
  title={COVID-19 risk perception framework of the public: an infodemic tool for future pandemics and epidemics},
  author={Lohiniva, Anna-Leena and Pensola, Annika and Hy{\"o}kki, Suvi and Sivel{\"a}, Jonas and Tammi, Tuukka},
  journal={BMC Public Health},
  volume={22},
  number={1},
  pages={2124},
  year={2022},
  publisher={Springer}
}

@article{ishizumi2024beyond,
  title={Beyond misinformation: developing a public health prevention framework for managing information ecosystems},
  author={Ishizumi, Atsuyoshi and Kolis, Jessica and Abad, Neetu and Prybylski, Dimitri and Brookmeyer, Kathryn A and Voegeli, Christopher and Wardle, Claire and Chiou, Howard},
  journal={The Lancet Public Health},
  volume={9},
  number={6},
  pages={e397--e406},
  year={2024},
  publisher={Elsevier}
}

@misc{crowdtangle2024,
  title={CrowdTangle},
  author={{CrowdTangle Team}},
  howpublished={Meta Transparency Center},
  year={2024},
  note={Meta, Menlo Park, California, United States}
}

@misc{globalrisks2024,
  title        = {The Global Risks Report 2024},
  author       = {{World Economic Forum}},
  year         = {2024},
  institution  = {World Economic Forum},
  address      = {Geneva, Switzerland},
  url          = {https://www.weforum.org/publications/global-risks-report-2024/},
  note         = {ISBN:978‑2‑940631‑64‑3}  
}

@book{WHO2024FalseInformationToolkit,
  title        = {Managing false information in health emergencies: an operational toolkit},
  author       = {{World Health Organization. Regional Office for Europe}},
  year         = {2024},
  publisher    = {World Health Organization. Regional Office for Europe},
  url          = {https://iris.who.int/handle/10665/375783},
  note         = {License: CC BY-NC-SA 3.0 IGO}
}

@book{WHO2025SocialListening,
  title        = {Social listening in infodemic management for public health emergencies: guidance on ethical considerations},
  author       = {{World Health Organization}},
  year         = {2025},
  publisher    = {World Health Organization},
  url          = {https://iris.who.int/handle/10665/381013},
  note         = {License: CC BY-NC-SA 3.0 IGO}
}

@article{cinelli2025infodemic,
  title={Infodemic Versus Viral Information Spread: Key Differences and Open Challenges},
  author={Cinelli, Matteo and Gesualdo, Francesco and others},
  journal={JMIR infodemiology},
  volume={5},
  number={1},
  pages={e57455},
  year={2025},
  publisher={JMIR Publications Inc., Toronto, Canada}
}

@article{sangiorgio2025evaluating,
  title={Evaluating the effect of viral posts on social media engagement},
  author={Sangiorgio, Emanuele and Di Marco, Niccol{\`o} and Etta, Gabriele and Cinelli, Matteo and Cerqueti, Roy and Quattrociocchi, Walter},
  journal={Scientific Reports},
  volume={15},
  number={1},
  pages={639},
  year={2025},
  publisher={Nature Publishing Group UK London}
}

@misc{WHO_GACVS_2021,
  author = {{World Health Organization}},
  title = {Global Advisory Committee on Vaccine Safety (GACVS) review of latest evidence of rare adverse blood coagulation events with AstraZeneca COVID-19 Vaccine (Vaxzevria and Covishield)},
  year  = {2021},
  month = apr # " 16",
  howpublished = {\url{https://www.who.int/news}}
}

@article{webster2021covid,
  title={COVID-19 timeline of events},
  author={Webster, Paul},
  journal={Nature medicine},
  volume={27},
  number={12},
  pages={2054--2055},
  year={2021},
  publisher={Nature Publishing Group US New York}
}

@misc{EMA_Moderna_2021,
  author       = {{European Medicines Agency}},
  title        = {EMA recommends COVID‑19 Vaccine Moderna for authorisation in the EU},
  year         = {2021},
  month        = jan,
  day          = 06,
  howpublished = {\url{https://www.ema.europa.eu/en/news/ema-recommends-covid-19-vaccine-moderna-authorisation-eu}}
}

@article{massicotte2016package,
  title={Package ‘gtrendsR’},
  author={Massicotte, Philippe and Eddelbuettel, Dirk and Massicotte, Maintainer Philippe},
  journal={R package},
  year={2016}
}

@article{piltch2023were,
  title={What were the information voids? A qualitative analysis of questions asked by Dear Pandemic readers between August 2020-August 2021},
  author={Piltch-Loeb, Rachael and James, Richard and Albrecht, Sandra S and Buttenheim, Alison M and Dowd, Jennifer BEAM and Kumar, Aparna and Jones, Malia and Leininger, Lindsey J and Simanek, Amanda and Aronowitz, Shoshana},
  journal={Journal of Health Communication},
  volume={28},
  number={sup1},
  pages={25--33},
  year={2023},
  publisher={Taylor \& Francis}
}

@article{dancho2020package,
  title={Package ‘anomalize’},
  author={Dancho, Matt and Vaughan, Davis},
  journal={R package},
  year={2020}
}

@article{cerqueti2024investors,
  title={Investors’ attention and network spillover for commodity market forecasting},
  author={Cerqueti, Roy and Ficcadenti, Valerio and Mattera, Raffaele},
  journal={Socio-Economic Planning Sciences},
  volume={95},
  pages={102023},
  year={2024},
  publisher={Elsevier}
}

@article{cleveland1990stl,
  title={STL: A seasonal-trend decomposition},
  author={Cleveland, Robert B and Cleveland, William S and McRae, Jean E and Terpenning, Irma and others},
  journal={J. off. Stat},
  volume={6},
  number={1},
  pages={3--73},
  year={1990}
}

@article{theodosiou2011forecasting,
  title={Forecasting monthly and quarterly time series using STL decomposition},
  author={Theodosiou, Marina},
  journal={International Journal of Forecasting},
  volume={27},
  number={4},
  pages={1178--1195},
  year={2011},
  publisher={Elsevier}
}

@article{pennycook2021psychology,
  title={The psychology of fake news},
  author={Pennycook, Gordon and Rand, David G},
  journal={Trends in cognitive sciences},
  volume={25},
  number={5},
  pages={388--402},
  year={2021},
  publisher={Elsevier}
}

@article{valensise2021entropy,
  title={Entropy and complexity unveil the landscape of memes evolution},
  author={Valensise, Carlo M and Serra, Alessandra and Galeazzi, Alessandro and Etta, Gabriele and Cinelli, Matteo and Quattrociocchi, Walter},
  journal={Scientific Reports},
  volume={11},
  number={1},
  pages={20022},
  year={2021},
  publisher={Nature Publishing Group UK London}
}

@article{cerqueti2024portfolio,
  title={Portfolio decision analysis for pandemic sentiment assessment based on finance and web queries},
  author={Cerqueti, Roy and Cesarone, Francesco and Ficcadenti, Valerio},
  journal={Annals of Operations Research},
  pages={1--31},
  year={2024},
  publisher={Springer}
}

@article{cerqueti2024anxiety,
  title={Anxiety about the pandemic and trust in financial markets},
  author={Cerqueti, Roy and Ficcadenti, Valerio},
  journal={The Annals of Regional Science},
  volume={72},
  number={4},
  pages={1277--1328},
  year={2024},
  publisher={Springer}
}

@article{budak2024misunderstanding,
  title={Misunderstanding the harms of online misinformation},
  author={Budak, Ceren and Nyhan, Brendan and Rothschild, David M and Thorson, Emily and Watts, Duncan J},
  journal={Nature},
  volume={630},
  number={8015},
  pages={45--53},
  year={2024},
  publisher={Nature Publishing Group UK London}
}

@article{veil2008cerc,
  title={CERC as a theoretical framework for research and practice},
  author={Veil, Shari and Reynolds, Barbara and Sellnow, Timothy L and Seeger, Matthew W},
  journal={Health promotion practice},
  volume={9},
  number={4\_suppl},
  pages={26S--34S},
  year={2008},
  publisher={Sage Publications Sage CA: Los Angeles, CA}
}

@techreport{world2022advancing,
  title={Advancing infodemic management in risk communication and community engagement in the WHO European Region: implementation guidance},
  author={World Health Organization and others},
  year={2022},
  institution={World Health Organization. Regional Office for Europe}
}

@techreport{ECDC2025_preparedness,
  author       = {{European Centre for Disease Prevention and Control}},
  title        = {Recommendations for preparedness planning for public health threats},
  institution  = {European Centre for Disease Prevention and Control},
  address      = {Stockholm},
  year         = {2025},
  doi          = {10.2900/7416543}
}

@article{wouters2022risk,
  title={Risk of death in nursing home residents after COVID-19 vaccination},
  author={Wouters, Fenne and van Loon, Anouk M and Rutten, Jeanine JS and Smalbrugge, Martin and Hertogh, Cees MPM and Joling, Karlijn J},
  journal={Journal of the American Medical Directors Association},
  volume={23},
  number={10},
  pages={1750--1753},
  year={2022},
  publisher={Elsevier}
}

@misc{FDA2021JanssenPauseLifted,
  author       = {{U.S. Food and Drug Administration}},
  title        = {FDA and CDC lift recommended pause on Johnson \& Johnson (Janssen) COVID-19 vaccine use following thorough safety review},
  year         = {2021},
  howpublished = {\url{https://www.fda.gov/news-events/press-announcements/fda-and-cdc-lift-recommended-pause-johnson-johnson-janssen-covid-19-vaccine-use-following-thorough}}
}

\clearpage

\part*{Supplementary Information}
\setcounter{section}{0}
\setcounter{figure}{0}
\renewcommand{\thefigure}{S\arabic{figure}}
\renewcommand{\thetable}{S\arabic{table}}

\section*{S1: Keyword search}
The following tables display the keyword sets used to collect vaccine-related data from major digital platforms: Facebook, Twitter, Google Trends, and Wikipedia. These keywords were selected to capture public discourse, search behaviour, and information access related to different COVID-19 vaccine brands across multiple countries.
Table~\ref{tab:facebook_twitter_gdelt} lists the keywords used to extract vaccine-related content from Facebook, Twitter, and GDELT. GDELT data were retrieved using the \texttt{GDELT summary} function, which supports keyword-based searches, using the same keywords as for the social media platforms. 
Table~\ref{tab:google_trends} reports the keywords used to monitor search interest via Google Trends, collected with the \texttt{gtrendsR} package in R \cite{massicotte2016package}. 
Table~\ref{tab:wikipedia} lists the keywords applied in Wikipedia page view analyses, specifying the countries in which each term was used. Notably, Google Trends queries were standardized across countries, whereas Wikipedia keywords were adapted to national search patterns and data availability.
\begin{table}[ht]
\centering
\small
\begin{tabular}{|l|p{10cm}|}
\hline
\textbf{Brand} & \textbf{Keywords} \\
\hline
AstraZeneca & Astrazeneca, astrazeneca, AstraZeneca, AZD1222, ChAdOx1, ChAdOx1 nCoV-2019, Vaxzevria \\
\hline
Moderna & mRNA-1273, moderna, Moderna \\
\hline
Pfizer & pfizer, biontech, pfizer/BioNTech, BNT16b2, Pfizer, Pfizer/BioNTech, pfizer/biontech \\
\hline
J\&J & JNJ-78436735, Johnson \& Johnson, J\&J, J\&j, j and j, johnson and johnson, Johnson and Johnson, Janssen \\
\hline
Sputnik & sputnik, Sputnik V, sputnik v, Gam-COVID-Vac, Gam-COVID-Vac Sputnik V, SputnikV, Gamaleya, Gamaleya Scientific Research Institute \\
\hline
\end{tabular}
\caption{Facebook, GDELT and Twitter-related vaccine keywords for each brand.}
\label{tab:facebook_twitter_gdelt}
\end{table}
\begin{table}[ht]
\centering
\small
\begin{tabular}{|l|p{9cm}|}
\hline
\textbf{Topic} & \textbf{Keywords} \\
\hline
AstraZeneca & AstraZeneca, AZD1222 \\
\hline
mRNA-1273 &mRNA-1273, moderna \\
\hline
Pfizer & pfizer, biontech \\
\hline
Johnson \& Johnson & JNJ-78436735, Johnson \& Johnson \\
\hline
Sputnik V & sputnik v, Gam-COVID-Vac \\
\hline
\end{tabular}
\caption{Google Trends vaccine-related keywords, used uniformly across all countries.}
\label{tab:google_trends}
\end{table}
\begin{table}[ht]
\centering
\small
\begin{tabular}{|l|p{8cm}|}
\hline
\textbf{Keyword} & \textbf{Country} \\
\hline
AstraZeneca & All countries \\
\hline
mRNA-1273 & Denmark, France, Germany, Italy, Spain \\
\hline
Moderna COVID-19 vaccine & United Kingdom \\
\hline
Pfizer & All countries \\
\hline
Johnson \& Johnson & All countries \\
\hline
Sputnik V & France, Germany, Italy, Spain, United Kingdom; no data for Sputnik V in Denmark \\
\hline
\end{tabular}
\caption{Wikipedia vaccine-related keywords and the countries in which they were used.}
\label{tab:wikipedia}
\end{table}

\section*{S2: Pipeline evaluation using synthetic data}
As shown in Table \ref{zscore}, standardized supply values in the empirical data reach Z-scores well above 30$\sigma$, motivating the choice of an upper bound of 15$\sigma$ in the anomaly simulations. The table reports the five largest observed Z-scores computed from the supply variable, illustrating the presence of extreme observations in the empirical data.
\begin{table}[ht]
\centering
\label{tab:si_extreme_supply}
\begin{tabular}{|l|l|l|c|c|l|}
\hline
Source & Date & Vaccine & Country & Supply & Z-score ($\sigma$) \\
\hline
Twitter  & 2020-11-16 & mRNA-1273         & DK & 83.54 & 37.69 \\
Twitter  & 2021-03-15 & AstraZeneca       & DE & 52.39 & 23.47 \\
Twitter  & 2021-04-13 & Johnson \& Johnson & UK & 40.82 & 18.18 \\
Twitter  & 2020-11-16 & Pfizer            & DK & 40.29 & 17.94 \\
Twitter  & 2021-03-15 & AstraZeneca       & FR & 38.54 & 17.14 \\
GDELT    & 2021-04-15 & Sputnik V         & DK & 57.06 & 27.23 \\
GDELT    & 2021-04-18 & Sputnik V         & DK & 57.06 & 27.23 \\
GDELT    & 2021-04-25 & Sputnik V         & DK & 57.06 & 27.23 \\
GDELT    & 2021-03-06 & mRNA-1273         & DK & 30.31 & 14.24 \\
GDELT    & 2021-04-24 & mRNA-1273         & DK & 30.31 & 14.24 \\
Facebook & 2021-01-08 & mRNA-1273         & UK & 47.84 & 22.10 \\
Facebook & 2021-04-14 & AstraZeneca       & DK & 47.70 & 22.03 \\
Facebook & 2021-03-11 & AstraZeneca       & DK & 45.60 & 21.04 \\
Facebook & 2020-12-02 & Pfizer            & UK & 43.89 & 20.24 \\
Facebook & 2021-04-07 & mRNA-1273         & UK & 40.93 & 18.84 \\
\hline
\end{tabular}
\caption{The table reports the largest observed Z-scores ($\sigma$) computed from the supply variable.}
\label{zscore}
\end{table}
Table~\ref{synt_data_results} reports the average precision and F1 score as a function of $\sigma$, obtained from simulations on synthetic data. Precision is defined as
\[
\text{Precision} = \frac{n_{\text{intersection}}}{n_{\text{detected}}},
\]
where $n_{\text{detected}}$ is the number of anomalies identified by the pipeline, and $n_{\text{intersection}}$ is the number of correctly detected anomalies.

The F1 score is computed as
\[
\text{F1} = \frac{2 \cdot \text{Precision} \cdot \text{Recall}}{\text{Precision} + \text{Recall}},
\]
with recall defined as
\[
\text{Recall} = \frac{n_{\text{intersection}}}{n_{\text{ground truth}}} \,
\]
where $n_{\text{ground truth}}$ measures the number of anomalies artificially introduced in the data.
Precision measures the proportion of detected anomalies that are true positives, reflecting the ability of the method to limit false positives, while the F1 score balances precision and recall to provide an overall assessment of detection performance. Synthetic scenarios are designed to mimic conditions analogous to information voids and overabundance, where imbalances between information demand and supply are particularly pronounced.
\begin{table}[ht]
\centering
\begin{tabular}{|l|l|l|}
\hline
${\sigma}$ & \textbf{Mean Precision} & \textbf{Mean F1 Score} \\
    \hline
    1.0 & 0.0 & 0.000 \\
    1.5 & 0.0 & 0.000 \\
    2.0 & 0.0 & 0.000 \\
    2.5 & 0.0 & 0.000 \\
    3.0 & 0.1 & 0.005 \\
    3.5 & 0.2 & 0.012 \\
    4.0 & 0.2 & 0.020 \\
    4.5 & 0.3 & 0.024 \\
    5.0 & 0.5 & 0.066 \\
    5.5 & 0.7 & 0.123 \\
    6.0 & 0.9 & 0.129 \\
    6.5 & 0.9 & 0.337 \\
    7.0 & 1.0 & 0.384 \\
    7.5 & 1.0 & 0.473 \\
    8.0 & 1.0 & 0.604 \\
    8.5 & 1.0 & 0.650 \\
    9.0 & 1.0 & 0.682 \\
    9.5 & 1.0 & 0.685 \\
    10.0 & 1.0 & 0.688 \\
    10.5 & 1.0 & 0.691 \\
    11.0 & 1.0 & 0.689 \\
    11.5 & 1.0 & 0.685 \\
    12.0 & 1.0 & 0.691 \\
    12.5 & 1.0 & 0.691 \\
    13.0 & 1.0 & 0.686 \\
    13.5 & 1.0 & 0.686 \\
    14.0 & 1.0 & 0.682 \\
    14.5 & 1.0 & 0.697 \\
    15.0 & 1.0 & 0.693 \\
\hline
\end{tabular}
\caption{Performance metrics for different values of $\sigma$.}
\label{synt_data_results}
\end{table}

\clearpage

\section*{S3: Time series visualization and number of pre and post-rollout observations}
\begin{figure}[H]
    \centering
    \includegraphics[width=1\linewidth]{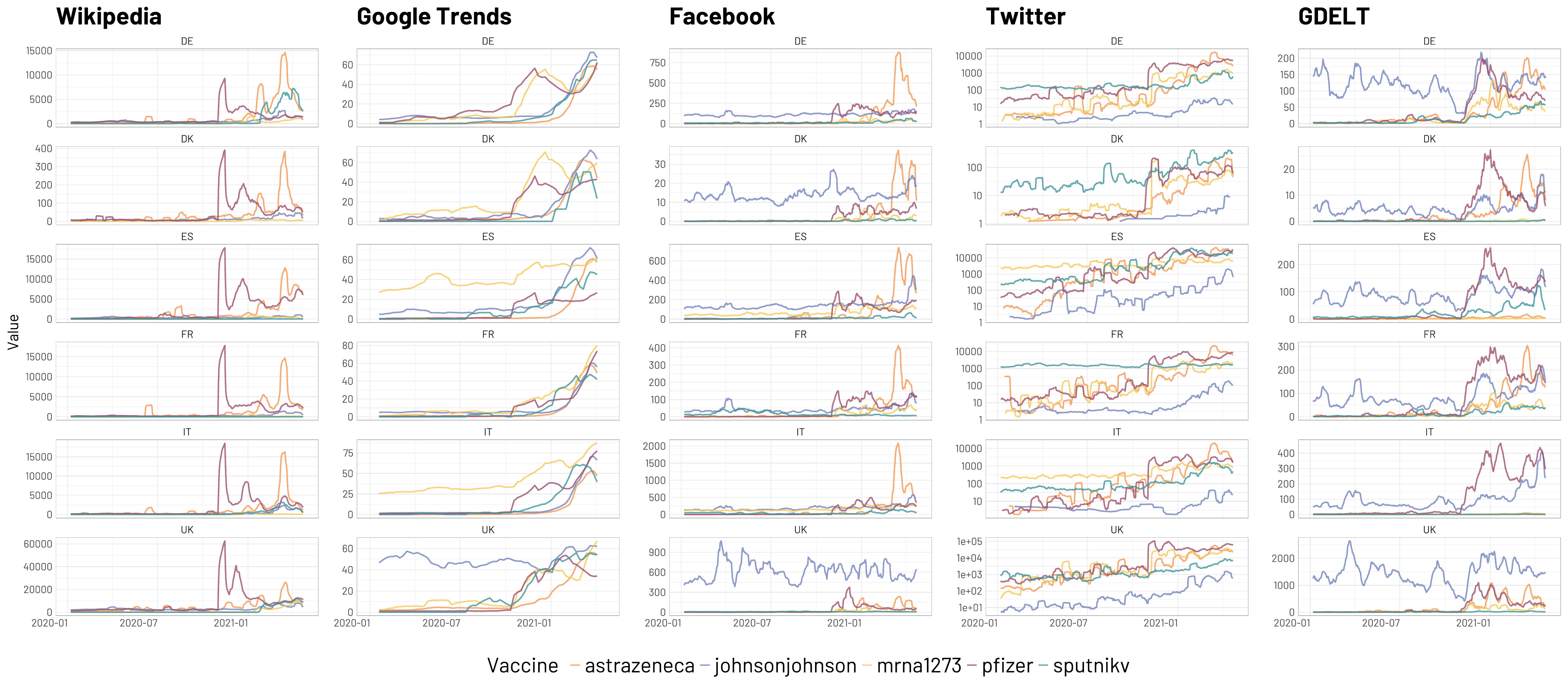}
    \caption{Time series of demand proxies - Wikipedia and Google Trends - and supply proxies - Facebook, Twitter, and GDELT - for the analysis period from January 1, 2020, to April 30, 2021. The series were smoothed using a 14-day rolling average for daily data and a 2-week rolling average for weekly data (Google Trends only).}
    \label{fig:ts}
\end{figure}

Figure~\ref{fig:ts} shows the temporal dynamics of demand and supply indicators related to COVID-19 vaccines over the period from 1 January 2020 to 30 April 2021. Demand is proxied by two measures of public information-seeking activity—Wikipedia page-views and Google Trends data—both capturing societal interest in vaccine-related topics. Supply is represented by content production across three major online platforms: Facebook, Twitter, and GDELT. To improve readability and reduce short-term noise, all time series were smoothed: daily datasets (Wikipedia, Facebook, Twitter, and GDELT) using a 14-day rolling average, and weekly Google Trends data using a 2-weeks rolling average. This figure allows a direct visual comparison of how public demand and supply evolved across platforms and throughout the pandemic.

We first analyse the temporal trajectories of these indicators. Although Google Trends provides weekly data while the other sources offer daily observations, all series exhibit a broadly increasing trend over the study period. Sharp peaks in both content production (GDELT articles, Twitter posts, Facebook posts) and information-seeking behaviour (Google searches, Wikipedia page-views) align with major pandemic milestones. Across countries, these temporal patterns are visually synchronized, highlighting a correspondence between public attention and media output during the most critical phases of the pandemic.

\clearpage
Table~\ref{tab:supply_data} reports the number of observations in the pre- and post-rollout periods for each information source, disaggregated by country and vaccine.
Table~\ref{tab:demand_data} reports the number of observations before and after the rollout; for some Wikipedia entries, all observations are recorded in the post-rollout period, as several pages were created shortly before or at the time of the vaccine rollout (e.g., Sputnik V, mrna1273 (Moderna)).

{
\setlength{\tabcolsep}{4pt}   
\renewcommand{\arraystretch}{0.95} 
\footnotesize

\begin{longtable}{|l|l|c|c|c|c|c|}
\caption{Number of observations before and after the vaccine rollout (27 December 2020), by source, vaccine, and country.}
\label{tab:supply_data}\\
\hline
Source & Vaccine & Country & Pre-rollout posts & Post-rollout posts & Pre-rollout \% & Post-rollout \%\\
\hline
\endfirsthead
\hline
Source & Vaccine & Country & Pre-rollout posts & Post-rollout posts & Pre-rollout \% & Post-rollout \%\\
\hline
\endhead
\hline
\hline
\endfoot
\hline
\endlastfoot
Facebook & AstraZeneca & DE & 1019 & 36949 & 2.7 & 97.3\\ \hline
Facebook & AstraZeneca & DK & 35 & 1585 & 2.2 & 97.8\\ \hline
Facebook & AstraZeneca & ES & 3264 & 34519 & 8.6 & 91.4\\ \hline
Facebook & AstraZeneca & FR & 1003 & 16023 & 5.9 & 94.1\\ \hline
Facebook & AstraZeneca & UK & 2203 & 14463 & 13.2 & 86.8\\ \hline
Facebook & AstraZeneca & IT & 4393 & 66892 & 6.2 & 93.8\\ \hline
Facebook & J\&J & DE & 38613 & 15947 & 70.8 & 29.2\\ \hline
Facebook & J\&J  & DK & 4984 & 1827 & 73.2 & 26.8\\ \hline
Facebook & J\&J  & ES & 46270 & 24225 & 65.6 & 34.4\\ \hline
Facebook & J\&J  & FR & 13493 & 7364 & 64.7 & 35.3\\ \hline
Facebook & J\&J  & UK & 223113 & 75789 & 74.6 & 25.4\\ \hline
Facebook & J\&J  & IT & 60063 & 34454 & 63.5 & 36.5\\ \hline
Facebook & mrna1273 & DE & 1856 & 5074 & 26.8 & 73.2\\ \hline
Facebook & mrna1273 & DK & 96 & 323 & 22.9 & 77.1\\ \hline
Facebook & mrna1273 & ES & 19800 & 13967 & 58.6 & 41.4\\ \hline
Facebook & mrna1273 & FR & 2090 & 5174 & 28.8 & 71.2\\ \hline
Facebook & mrna1273 & UK & 1645 & 3140 & 34.4 & 65.6\\ \hline
Facebook & mrna1273 & IT & 47683 & 30073 & 61.3 & 38.7\\ \hline
Facebook & Pfizer & DE & 10764 & 18658 & 36.6 & 63.4\\ \hline
Facebook & Pfizer & DK & 328 & 703 & 31.8 & 68.2\\ \hline
Facebook & Pfizer & ES & 9781 & 16896 & 36.7 & 63.3\\ \hline
Facebook & Pfizer & FR & 5580 & 11307 & 33.0 & 67.0\\ \hline
Facebook & Pfizer & UK & 9300 & 8094 & 53.5 & 46.5\\ \hline
Facebook & Pfizer & IT & 12329 & 33145 & 27.1 & 72.9\\ \hline
Facebook & Sputnik V & DE & 2918 & 3598 & 44.8 & 55.2\\ \hline
Facebook & Sputnik V & DK & 53 & 86 & 38.1 & 61.9\\ \hline
Facebook & Sputnik V & ES & 2511 & 2797 & 47.3 & 52.7\\ \hline
Facebook & Sputnik V & FR & 7022 & 1042 & 87.1 & 12.9\\ \hline
Facebook & Sputnik V & UK & 2642 & 919 & 74.2 & 25.8\\ \hline
Facebook & Sputnik V & IT & 11070 & 9670 & 53.4 & 46.6\\ \hline
GDELT & AstraZeneca & DE & 3054 & 14499 & 17.4 & 82.6\\ \hline
GDELT & AstraZeneca & DK & 224 & 1341 & 14.3 & 85.7\\ \hline
GDELT & AstraZeneca & ES & 287 & 1032 & 21.8 & 78.2\\ \hline
GDELT & AstraZeneca & FR & 3219 & 17421 & 15.6 & 84.4\\ \hline
GDELT & AstraZeneca & UK & 20009 & 73213 & 21.5 & 78.5\\ \hline
GDELT & AstraZeneca & IT & 340 & 561 & 37.7 & 62.3\\ \hline
GDELT & J\&J  & DE & 42906 & 16254 & 72.5 & 27.5\\ \hline
GDELT & J\&J  & DK & 1567 & 939 & 62.5 & 37.5\\ \hline
GDELT & J\&J  & ES & 29163 & 14304 & 67.1 & 32.9\\ \hline
GDELT & J\&J  & FR & 26123 & 15512 & 62.7 & 37.3\\ \hline
GDELT & J\&J  & UK & 494201 & 209126 & 70.3 & 29.7\\ \hline
GDELT & J\&J  & IT & 25475 & 21986 & 53.7 & 46.3\\ \hline
GDELT & mrna1273 & DE & 3700 & 7645 & 32.6 & 67.4\\ \hline
GDELT & mrna1273 & DK & 6 & 42 & 12.5 & 87.5\\ \hline
GDELT & mrna1273 & ES & 122 & 253 & 32.5 & 67.5\\ \hline
GDELT & mrna1273 & FR & 2920 & 6823 & 30.0 & 70.0\\ \hline
GDELT & mrna1273 & UK & 10481 & 20153 & 34.2 & 65.8\\ \hline
GDELT & mrna1273 & IT & 51 & 134 & 27.6 & 72.4\\ \hline
GDELT & Pfizer & DE & 7929 & 12312 & 39.2 & 60.8\\ \hline
GDELT & Pfizer & DK & 637 & 1444 & 30.6 & 69.4\\ \hline
GDELT & Pfizer & ES & 7471 & 18460 & 28.8 & 71.2\\ \hline
GDELT & Pfizer & FR & 11355 & 24644 & 31.5 & 68.5\\ \hline
GDELT & Pfizer & UK & 32668 & 54767 & 37.4 & 62.6\\ \hline
GDELT & Pfizer & IT & 13606 & 41576 & 24.7 & 75.3\\ \hline
GDELT & Sputnik V & DE & 2312 & 5040 & 31.4 & 68.6\\ \hline
GDELT & Sputnik V & DK & 3 & 14 & 17.6 & 82.4\\ \hline
GDELT & Sputnik V & ES & 4494 & 6310 & 41.6 & 58.4\\ \hline
GDELT & Sputnik V & FR & 2260 & 4584 & 33.0 & 67.0\\ \hline
GDELT & Sputnik V & UK & 2959 & 2991 & 49.7 & 50.3\\ \hline
GDELT & Sputnik V & IT & 22 & 108 & 16.9 & 83.1\\ \hline
Twitter & AstraZeneca & DE & 10951 & 554804 & 1.9 & 98.1\\ \hline
Twitter & AstraZeneca & DK & 386 & 9617 & 3.9 & 96.1\\ \hline
Twitter & AstraZeneca & ES & 445991 & 2224880 & 16.7 & 83.3\\ \hline
Twitter & AstraZeneca & FR & 39700 & 763071 & 4.9 & 95.1\\ \hline
Twitter & AstraZeneca & UK & 584446 & 2854080 & 17.0 & 83.0\\ \hline
Twitter & AstraZeneca & IT & 22860 & 535767 & 4.1 & 95.9\\ \hline
Twitter & J\&J  & DE & 342 & 2178 & 13.6 & 86.4\\ \hline
Twitter & J\&J  & DK & 13 & 257 & 4.8 & 95.2\\ \hline
Twitter & J\&J  & ES & 7910 & 55590 & 12.5 & 87.5\\ \hline
Twitter & J\&J  & FR & 480 & 7573 & 6.0 & 94.0\\ \hline
Twitter & J\&J  & UK & 9667 & 71476 & 11.9 & 88.1\\ \hline
Twitter & J\&J  & IT & 273 & 1868 & 12.8 & 87.2\\ \hline
Twitter & mrna1273 & DE & 30167 & 107808 & 21.9 & 78.1\\ \hline
Twitter & mrna1273 & DK & 3934 & 5058 & 43.8 & 56.2\\ \hline
Twitter & mrna1273 & ES & 1333809 & 801967 & 62.5 & 37.5\\ \hline
Twitter & mrna1273 & FR & 65503 & 193090 & 25.3 & 74.7\\ \hline
Twitter & mrna1273 & UK & 1493734 & 2362883 & 38.7 & 61.3\\ \hline
Twitter & mrna1273 & IT & 114821 & 127142 & 47.5 & 52.5\\ \hline
Twitter & Pfizer & DE & 159560 & 500318 & 24.2 & 75.8\\ \hline
Twitter & Pfizer & DK & 5065 & 9381 & 35.1 & 64.9\\ \hline
Twitter & Pfizer & ES & 1251805 & 2751021 & 31.3 & 68.7\\ \hline
Twitter & Pfizer & FR & 204544 & 722592 & 22.1 & 77.9\\ \hline
Twitter & Pfizer & UK & 3892129 & 5071758 & 43.4 & 56.6\\ \hline
Twitter & Pfizer & IT & 106546 & 285559 & 27.2 & 72.8\\ \hline
Twitter & Sputnik V & DE & 54366 & 70708 & 43.5 & 56.5\\ \hline
Twitter & Sputnik V & DK & 13733 & 27403 & 33.4 & 66.6\\ \hline
Twitter & Sputnik V & ES & 712784 & 2734042 & 20.7 & 79.3\\ \hline
Twitter & Sputnik V & FR & 539975 & 207857 & 72.2 & 27.8\\ \hline
Twitter & Sputnik V & UK & 367319 & 523150 & 41.3 & 58.7\\ \hline
Twitter & Sputnik V & IT & 23290 & 85140 & 21.5 & 78.5\\ \hline
\end{longtable}
}

\clearpage
{
\setlength{\tabcolsep}{4pt}   
\renewcommand{\arraystretch}{0.95} 
\footnotesize

\begin{longtable}{|l|l|c|c|c|c|c|}
\caption{Number of observations before and after the vaccine rollout (27 December 2020), by source, vaccine, and country.}
\label{tab:demand_data}\\
\hline
Source & Vaccine & Country & Pre-rollout posts & Post-rollout posts & Pre-rollout \% & Post-rollout \%\\
\hline
\endfirsthead
\hline
Source & Vaccine & Country & Pre-rollout posts & Post-rollout posts & Pre-rollout \% & Post-rollout \%\\
\hline
\endhead
\hline
\multicolumn{7}{r}{\textit{Continued on next page}}\\
\hline
\endfoot
\hline
\endlastfoot
Wikipedia & AstraZeneca & DE & 121531 & 679364 & 15.2 & 84.8\\ \hline
Wikipedia & AstraZeneca & DK & 5144 & 13387 & 27.8 & 72.2\\ \hline
Wikipedia & AstraZeneca & ES & 189869 & 623497 & 23.3 & 76.7\\ \hline
Wikipedia & AstraZeneca & FR & 131100 & 504938 & 20.6 & 79.4\\ \hline
Wikipedia & AstraZeneca & UK & 938870 & 1372305 & 40.6 & 59.4\\ \hline
Wikipedia & AstraZeneca & IT & 114477 & 491291 & 18.9 & 81.1\\ \hline
Wikipedia & J\&J  & DE & 124718 & 151672 & 45.1 & 54.9\\ \hline
Wikipedia & J\&J  & DK & 2911 & 3191 & 47.7 & 52.3\\ \hline
Wikipedia & J\&J  & ES & 130939 & 70492 & 65.0 & 35.0\\ \hline
Wikipedia & J\&J  & FR & 72567 & 83829 & 46.4 & 53.6\\ \hline
Wikipedia & J\&J  & UK & 909860 & 613006 & 59.7 & 40.3\\ \hline
Wikipedia & J\&J  & IT & 42468 & 109935 & 27.9 & 72.1\\ \hline
Wikipedia & mrna1273 & DE & 9076 & 91439 & 9.0 & 91.0\\ \hline
Wikipedia & mrna1273 & DK & 48 & 867 & 5.2 & 94.8\\ \hline
Wikipedia & mrna1273 & ES & 0 & 60771 & 0.0 & 100.0\\ \hline
Wikipedia & mrna1273 & FR & 0 & 21419 & 0.0 & 100.0\\ \hline
Wikipedia & mrna1273 & UK & 26 & 531547 & 0.0 & 100.0\\ \hline
Wikipedia & mrna1273 & IT & 0 & 21465 & 0.0 & 100.0\\ \hline
Wikipedia & Pfizer & DE & 341587 & 205359 & 62.5 & 37.5\\ \hline
Wikipedia & Pfizer & DK & 12325 & 9170 & 57.3 & 42.7\\ \hline
Wikipedia & Pfizer & ES & 572756 & 611606 & 48.4 & 51.6\\ \hline
Wikipedia & Pfizer & FR & 387276 & 313456 & 55.3 & 44.7\\ \hline
Wikipedia & Pfizer & UK & 2418165 & 1272028 & 65.5 & 34.5\\ \hline
Wikipedia & Pfizer & IT & 442481 & 395675 & 52.8 & 47.2\\ \hline
Wikipedia & Sputnik V & DE & 5107 & 402295 & 1.3 & 98.7\\ \hline
Wikipedia & Sputnik V & ES & 307 & 3036 & 9.2 & 90.8\\ \hline
Wikipedia & Sputnik V & FR & 0 & 455 & 0.0 & 100.0\\ \hline
Wikipedia & Sputnik V & UK & 0 & 778739 & 0.0 & 100.0\\ \hline
Wikipedia & Sputnik V & IT & 6793 & 163833 & 4.0 & 96.0\\ \hline
Google Trends & AstraZeneca & DE & 19 & 584 & 3.2 & 96.8\\ \hline
Google Trends & AstraZeneca & DK & 81 & 674 & 10.7 & 89.3\\ \hline
Google Trends & AstraZeneca & ES & 32 & 573 & 5.3 & 94.7\\ \hline
Google Trends & AstraZeneca & FR & 23 & 545 & 4.0 & 96.0\\ \hline
Google Trends & AstraZeneca & UK & 210 & 706 & 22.9 & 77.1\\ \hline
Google Trends & AstraZeneca & IT & 14 & 493 & 2.8 & 97.2\\ \hline
Google Trends & J\&J  & DE & 326 & 840 & 28.0 & 72.0\\ \hline
Google Trends & J\&J  & DK & 214 & 856 & 20.0 & 80.0\\ \hline
Google Trends & J\&J  & ES & 433 & 901 & 32.5 & 67.5\\ \hline
Google Trends & J\&J  & FR & 258 & 592 & 30.4 & 69.6\\ \hline
Google Trends & J\&J  & UK & 2465 & 1040 & 70.3 & 29.7\\ \hline
Google Trends & J\&J  & IT & 111 & 692 & 13.8 & 86.2\\ \hline
Google Trends & mrna1273 & DE & 545 & 869 & 38.5 & 61.5\\ \hline
Google Trends & mrna1273 & DK & 780 & 992 & 44.0 & 56.0\\ \hline
Google Trends & mrna1273 & ES & 1955 & 1023 & 65.6 & 34.4\\ \hline
Google Trends & mrna1273 & FR & 322 & 912 & 26.1 & 73.9\\ \hline
Google Trends & mrna1273 & UK & 576 & 858 & 40.2 & 59.8\\ \hline
Google Trends & mrna1273 & IT & 1727 & 1316 & 56.8 & 43.2\\ \hline
Google Trends & Pfizer & DE & 787 & 807 & 49.4 & 50.6\\ \hline
Google Trends & Pfizer & DK & 402 & 656 & 38.0 & 62.0\\ \hline
Google Trends & Pfizer & ES & 228 & 392 & 36.8 & 63.2\\ \hline
Google Trends & Pfizer & FR & 165 & 754 & 18.0 & 82.0\\ \hline
Google Trends & Pfizer & UK & 323 & 753 & 30.0 & 70.0\\ \hline
Google Trends & Pfizer & IT & 315 & 960 & 24.7 & 75.3\\ \hline
Google Trends & Sputnik V & DE & 70 & 733 & 8.7 & 91.3\\ \hline
Google Trends & Sputnik V & DK & 0 & 441 & 0.0 & 100.0\\ \hline
Google Trends & Sputnik V & ES & 168 & 651 & 20.5 & 79.5\\ \hline
Google Trends & Sputnik V & FR & 55 & 642 & 7.9 & 92.1\\ \hline
Google Trends & Sputnik V & UK & 366 & 917 & 28.5 & 71.5\\ \hline
Google Trends & Sputnik V & IT & 111 & 744 & 13.0 & 87.0\\ \hline
\end{longtable}
}

\clearpage

\section*{S4: Compute the delta as supply minus demand: Google Trends and Wikipedia}

\begin{figure}[H]
    \centering
    \includegraphics[width=1\linewidth]{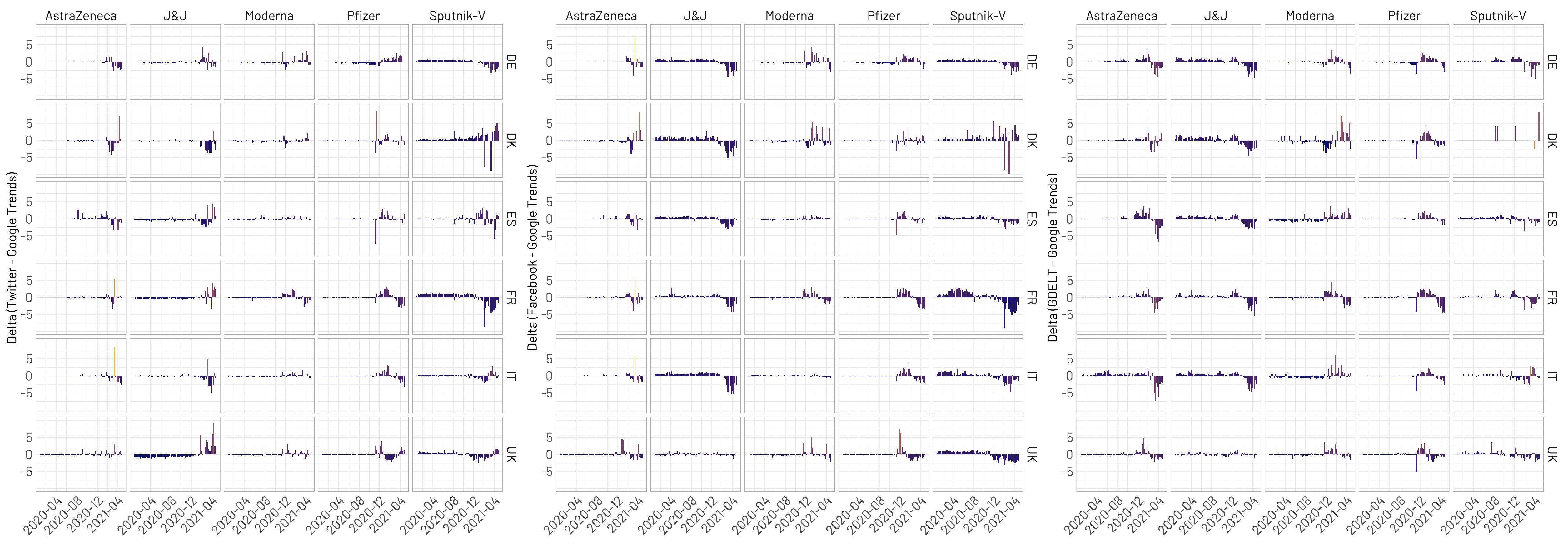}
    \caption{Weekly variation of the supply-demand delta across countries, vaccines, and sources. Colour intensity represents supply levels. Each panel displays trends over time for a specific combination of country, vaccine, and source.}
    \label{fig:delta_gt}
\end{figure}

\begin{figure}[H]
    \centering
    \includegraphics[width=1\linewidth]{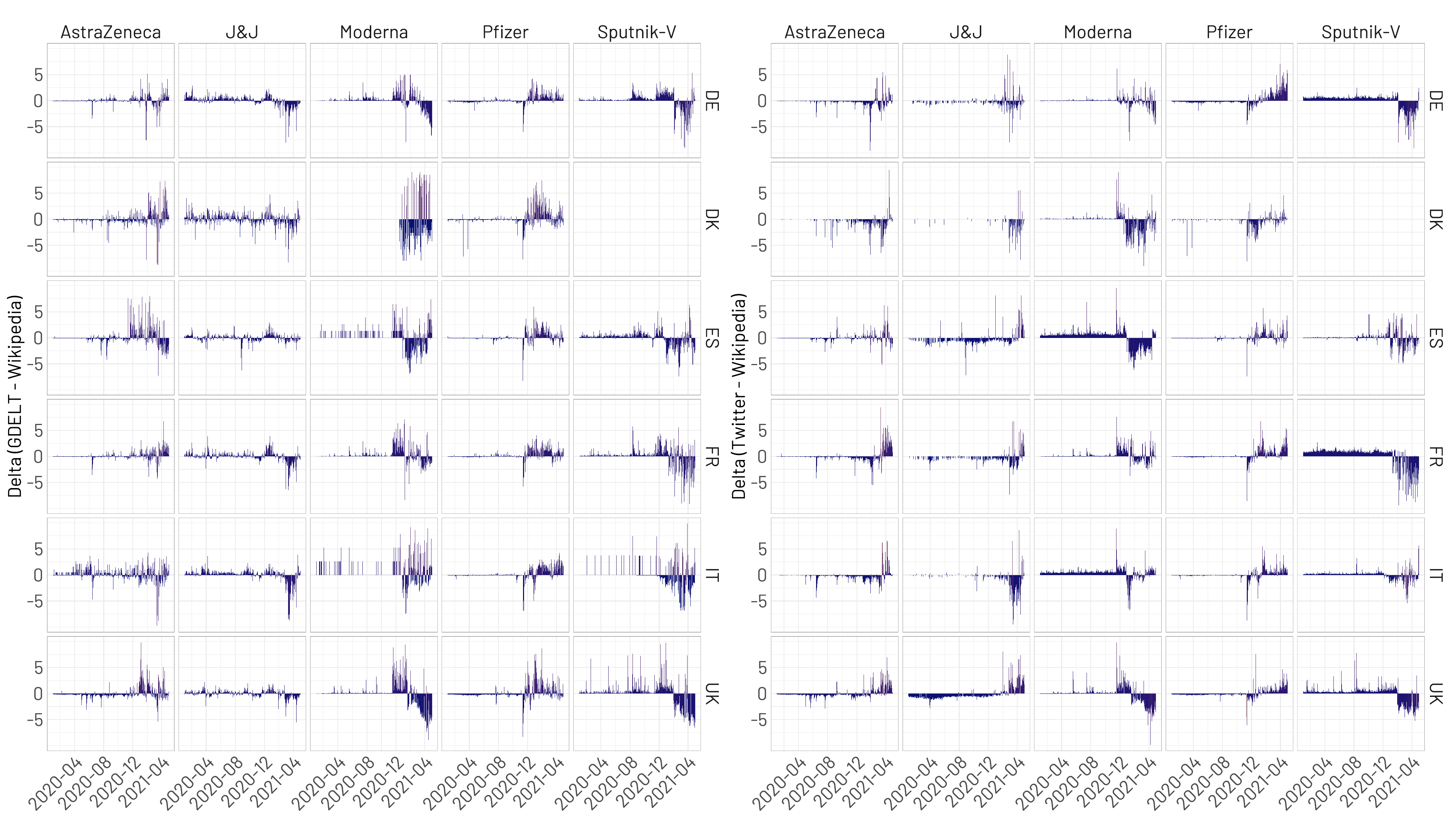}
    \caption{Daily variation of the supply-demand delta across countries, vaccines, and sources. Colour intensity represents supply levels. Each panel displays trends over time for a specific combination of country, vaccine, and source.}
    \label{fig:delta_wiki}
\end{figure}

Figure \ref{fig:delta_gt} presents results regarding the information delta computed using Google Trends, instead of Wikipedia, as the demand proxy. The resulting patterns are consistent with those obtained from the Wikipedia-based deltas, with the key difference that Google Trends data are available at a weekly rather than a daily resolution.
Figure ~\ref{fig:delta_wiki} shows the delta computed using Wikipedia page-views as the demand proxy and GDELT and Twitter activity as the supply proxies.

\section*{S5: Cross correlation result}
To further assess the responsiveness between information demand and supply, we computed the cross-correlation between their respective time series. Across most countries and vaccines, the highest correlations occur at or near lag 0, indicating a largely synchronous relationship between public interest and content production. The following tables report cross-correlation coefficients at lag 0 between vaccine-related demand and online information supply across platforms and countries. Table~\ref{tab:wiki_cross} uses Wikipedia page views as a proxy for demand, while Table~\ref{tab:gt_cross} relies on Google Trends data. In both cases, Facebook, Twitter, and GDELT are used as indicators of media supply. Correlation values are reported by vaccine brand and by country.

\begin{table}[ht]
\centering
\small
\begin{tabular}{|l|l|c|c|c|}
\hline
\textbf{Country} & \textbf{Vaccine} & \textbf{Facebook} & \textbf{Twitter} & \textbf{GDELT} \\
\hline
DE & AstraZeneca & 0.809 & 0.857 & 0.807 \\
DE & J\&J  & 0.532 & 0.577 & 0.182 \\
DE & mrna1273 & 0.728 & 0.702 & 0.615 \\
DE & Pfizer & 0.659 & 0.504 & 0.330 \\
DE & Sputnik V & 0.833 & 0.894 & 0.750 \\
DK & AstraZeneca & 0.676 & 0.653 & 0.799 \\
DK & J\&J  & 0.220 & 0.569 & 0.508 \\
DK & mrna1273 & 0.390 & 0.123 & 0.348 \\
DK & Pfizer & 0.448 & 0.297 & 0.376 \\
ES & AstraZeneca & 0.842 & 0.917 & 0.555 \\
ES & J\&J  & 0.545 & 0.500 & 0.370 \\
ES & mrna1273 & 0.483 & 0.381 & 0.316 \\
ES & Pfizer & 0.709 & 0.648 & 0.381 \\
ES & Sputnik V & 0.481 & 0.706 & 0.600 \\
FR & AstraZeneca & 0.828 & 0.874 & 0.746 \\
FR & J\&J  & 0.517 & 0.711 & 0.425 \\
FR & mrna1273 & 0.576 & 0.541 & 0.543 \\
FR & Pfizer & 0.558 & 0.455 & 0.244 \\
FR & Sputnik V & -0.169 & 0.215 & 0.682 \\
UK & AstraZeneca & 0.731 & 0.906 & 0.761 \\
UK & J\&J  & 0.002 & 0.792 & 0.128 \\
UK & mrna1273 & 0.322 & 0.511 & 0.425 \\
UK & Pfizer & 0.618 & 0.900 & 0.405 \\
UK & Sputnik V & 0.070 & 0.749 & 0.323 \\
IT & AstraZeneca & 0.879 & 0.946 & 0.662 \\
IT & J\&J  & 0.588 & 0.657 & 0.674 \\
IT & mrna1273 & 0.341 & 0.457 & 0.105 \\
IT & Pfizer & 0.599 & 0.706 & 0.363 \\
IT & Sputnik V & 0.677 & 0.911 & 0.690 \\
\hline
\end{tabular}
\caption{Cross-correlation at lag 0 between demand (Wikipedia) and supply (Facebook, Twitter, and GDELT) by country and vaccine.}
\label{tab:wiki_cross}
\end{table}

\begin{table}[ht]
\centering
\small
\begin{tabular}{|l|l|c|c|c|}
\hline
\textbf{Country} & \textbf{Vaccine} & \textbf{Facebook} & \textbf{Twitter} & \textbf{GDELT} \\
\hline
DE & AstraZeneca & 0.927 & 0.891 & 0.884 \\
DE & J\&J  & 0.579 & 0.850 & 0.165 \\
DE & mrna1273 & 0.446 & 0.705 & 0.402 \\
DE & Pfizer & 0.915 & 0.833 & 0.693 \\
DE & Sputnik V & 0.926 & 0.949 & 0.886 \\
DK & AstraZeneca & 0.850 & 0.789 & 0.878 \\
DK & J\&J  & 0.004 & 0.083 & -0.106 \\
DK & mrna1273 & 0.541 & 0.353 & 0.675 \\
DK & Pfizer & 0.839 & 0.695 & 0.674 \\
DK & Sputnik V & 0.630 & 0.759 & 0.535 \\
ES & AstraZeneca & 0.952 & 0.910 & 0.735 \\
ES & J\&J  & 0.624 & 0.687 & 0.476 \\
ES & mrna1273 & 0.621 & 0.467 & 0.721 \\
ES & Pfizer & 0.882 & 0.778 & 0.570 \\
ES & Sputnik V & 0.919 & 0.612 & 0.934 \\
FR & AstraZeneca & 0.926 & 0.948 & 0.915 \\
FR & J\&J  & -0.073 & -0.052 & -0.077 \\
FR & mrna1273 & 0.664 & 0.703 & 0.525 \\
FR & Pfizer & 0.756 & 0.797 & 0.571 \\
FR & Sputnik V & -0.339 & 0.267 & 0.876 \\
UK & AstraZeneca & 0.853 & 0.961 & 0.789 \\
UK & J\&J  & 0.000 & 0.930 & 0.177 \\
UK & mrna1273 & 0.880 & 0.910 & 0.862 \\
UK & Pfizer & 0.723 & 0.816 & 0.723 \\
UK & Sputnik V & 0.091 & 0.854 & 0.709 \\
IT & AstraZeneca & 0.949 & 0.930 & 0.869 \\
IT & J\&J  & 0.747 & 0.809 & 0.828 \\
IT & mrna1273 & 0.726 & 0.718 & 0.564 \\
IT & Pfizer & 0.817 & 0.831 & 0.832 \\
IT & Sputnik V & 0.759 & 0.964 & 0.883 \\
\hline
\end{tabular}
\caption{Cross-correlation at lag 0 between demand (Google Trends) and supply (Facebook, Twitter, and GDELT) by country and vaccine.}
\label{tab:gt_cross}
\end{table}

\clearpage

\section*{S6: Anomaly Detection}
\begin{figure}[ht]
    \centering
    \includegraphics[width=1\linewidth]{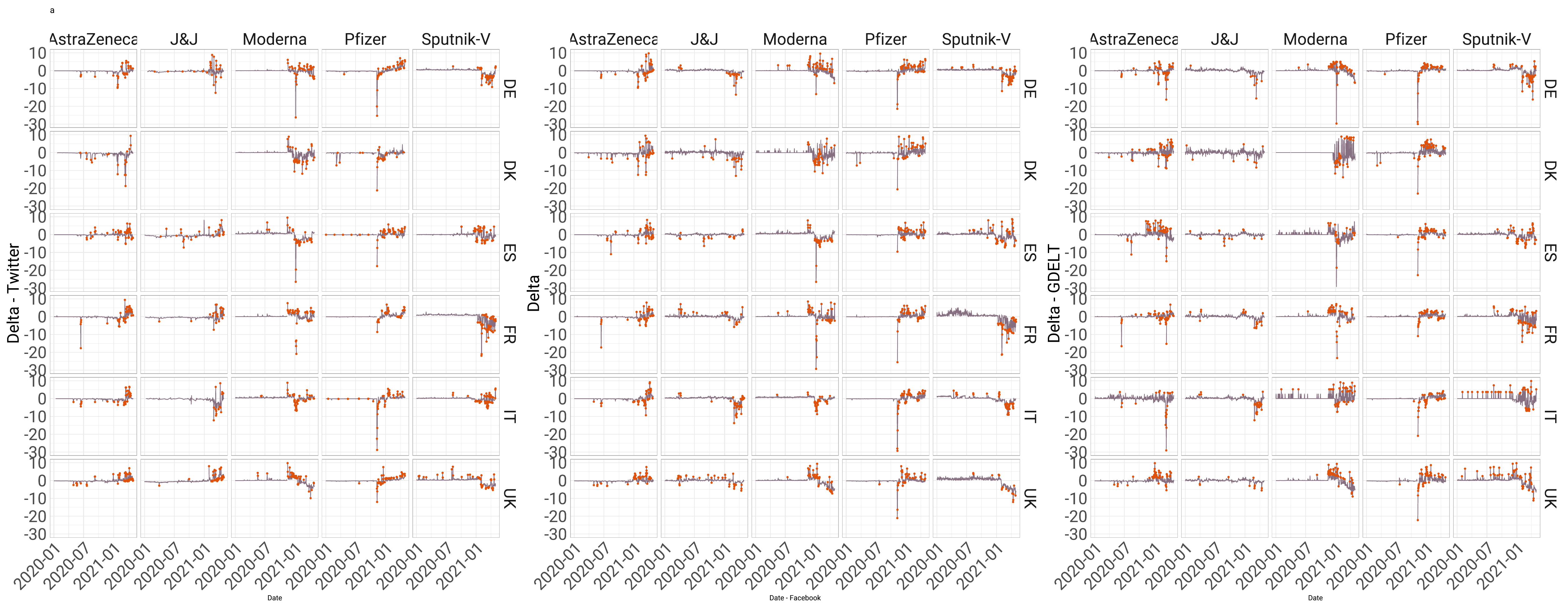}
    \caption{Result of the anomaly detection analysis, where Wikipedia serves as the demand proxy. The red points represent the anomalies flagged by the algorithm. We set $max\_anoms = 10\%$, which limits the maximum percentage of points that can be flagged as anomalies.}
    \label{fig:anom_det_wiki}
\end{figure}

\begin{figure}[ht]
    \centering
    \includegraphics[width=1\linewidth]{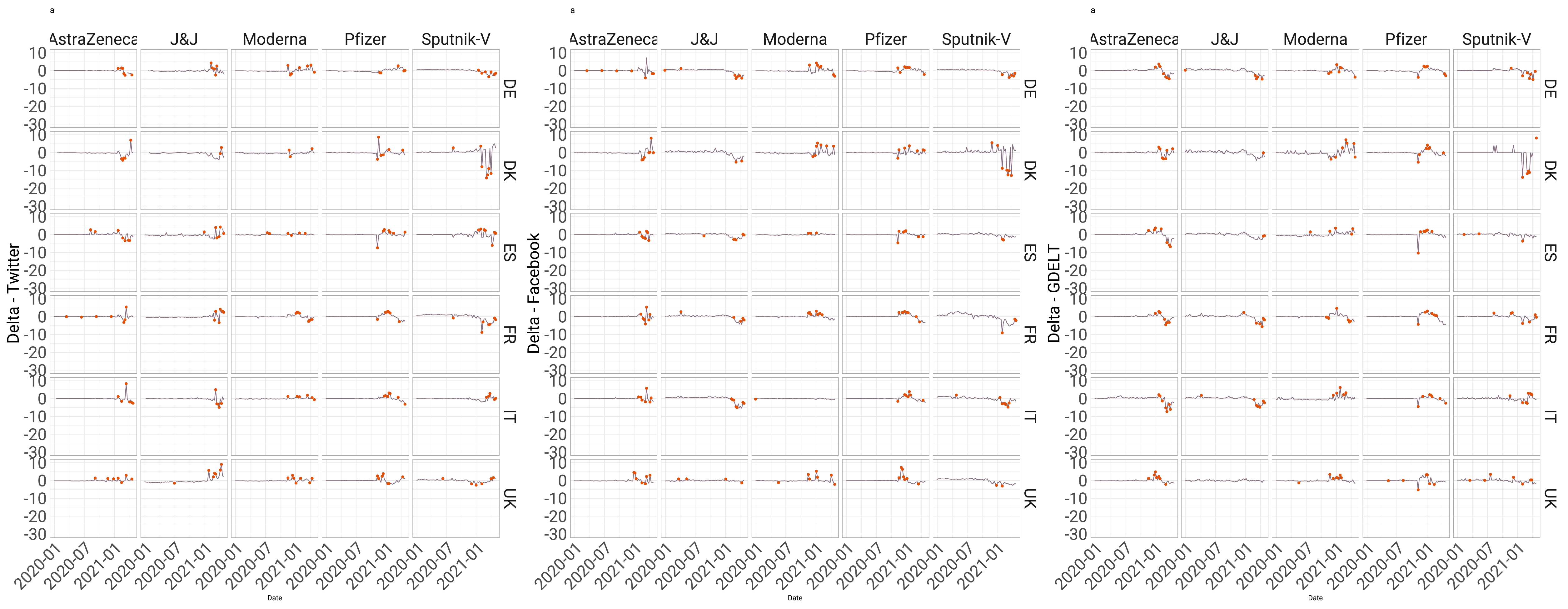}
    \caption{Result of the anomaly detection analysis, where Google Trends serves as the demand proxy. The red points represent the anomalies flagged by the algorithm. We set \texttt{max\_anom = 10\%}, which limits the maximum percentage of points that can be flagged as anomalies.}
    \label{fig:anom_det_gt}
\end{figure}

The anomaly detection process (Figures \ref{fig:anom_det_gt} and \ref{fig:anom_det_wiki}) follows a four-step procedure using a non-supervised approach \cite{dancho2020package}. The steps are as follows:

\subsection*{1. Time Decompose} 
$method = stl$, Seasonal-Trend decomposition using Loess. A seasonal decomposition is performed on delta, producing four components:
\begin{itemize}
    \item \textbf{Observed}: The observed values (delta).
    \item \textbf{Season}: The seasonal trend, with weekly seasonality for daily data.
    \item \textbf{Trend}: The long-term trend, obtained using a Loess smoother with a 3-month span for daily data.
    \item \textbf{Remainder}: The residual component, calculated as the difference between observed values and both season and trend: 
    \[
    \text{remainder} = \text{observed} - (\text{season} + \text{trend})
    \]
\end{itemize}

\subsection*{2. Anomalize ($method = iqr \,, alpha = 0.05 \,, max \_anoms = 0.10$)}
Anomaly detection is performed on the Remainder column. Where $alpha$ controls the width of the "normal" range. Lower values are more conservative while higher values are less prone to incorrectly classifying "normal" observations, and $max\_anoms$ is maximum percent of anomalies permitted to be identified. 
Three new columns are created:
\begin{itemize}
    \item \textbf{remainder\_l1}: lower limit of the remainder, calculated as: 
    \[
    \text{LL} = Q1 - k \times \text{IQR}
    \]
    \item \textbf{remainder\_l2}: upper limit of the remainder, calculated as: 
    \[
    \text{UL} = Q3 + k \times \text{IQR}
    \]
    \item \textbf{Anomaly}: a label indicating whether a value is an anomaly (Yes/No) based on the remainder exceeding the limits.
\end{itemize}

\subsection*{3. Time Recompose}
The seasonal, trend, and remainder components are recomposed into new limits that bound the observed values. The following columns are created:
\begin{itemize}
    \item \textbf{recomposed\_l1}: Lower bound of outliers around the observed value.
    \item \textbf{recomposed\_l2}: Upper bound of outliers around the observed value.
\end{itemize}

\subsection*{Non-Supervised Method}
The entire process is carried out using a non-supervised method. The algorithm autonomously detects anomalies based on the statistical structure of the time series data, using seasonal decomposition and IQR analysis.

\section*{S7: Anomaly Detection: Google Trends and Wikipedia results}
\begin{figure}[ht]
    \centering
    \includegraphics[width=1\linewidth]{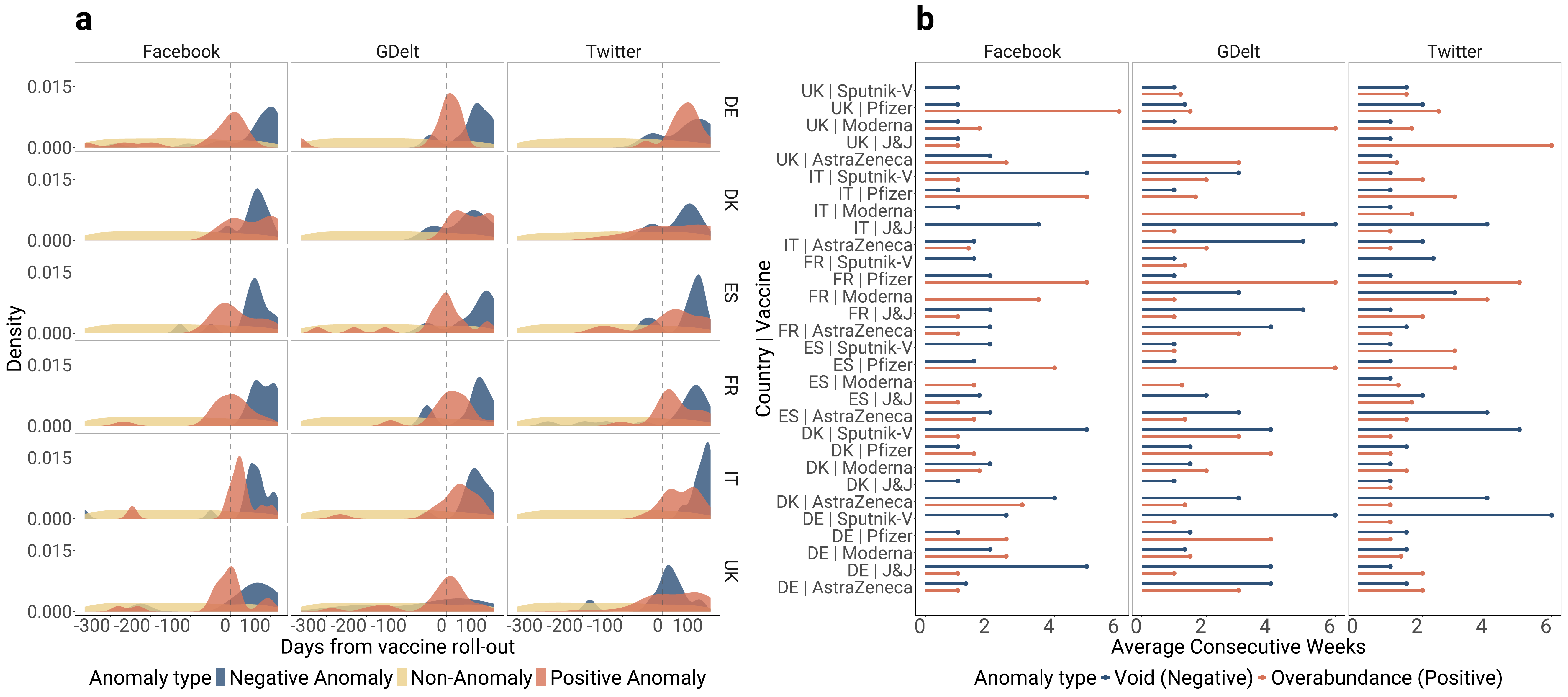}
    \caption{a) Distribution of anomalies before and after the vaccine roll-out, b) Average duration of Anomalies by Source Negative anomalies (blue) vs Positive anomalies (red)}
    \label{fig:gt_anomaly_det}
\end{figure}

\begin{figure}[H]
    \centering
    \includegraphics[width=0.8\linewidth]{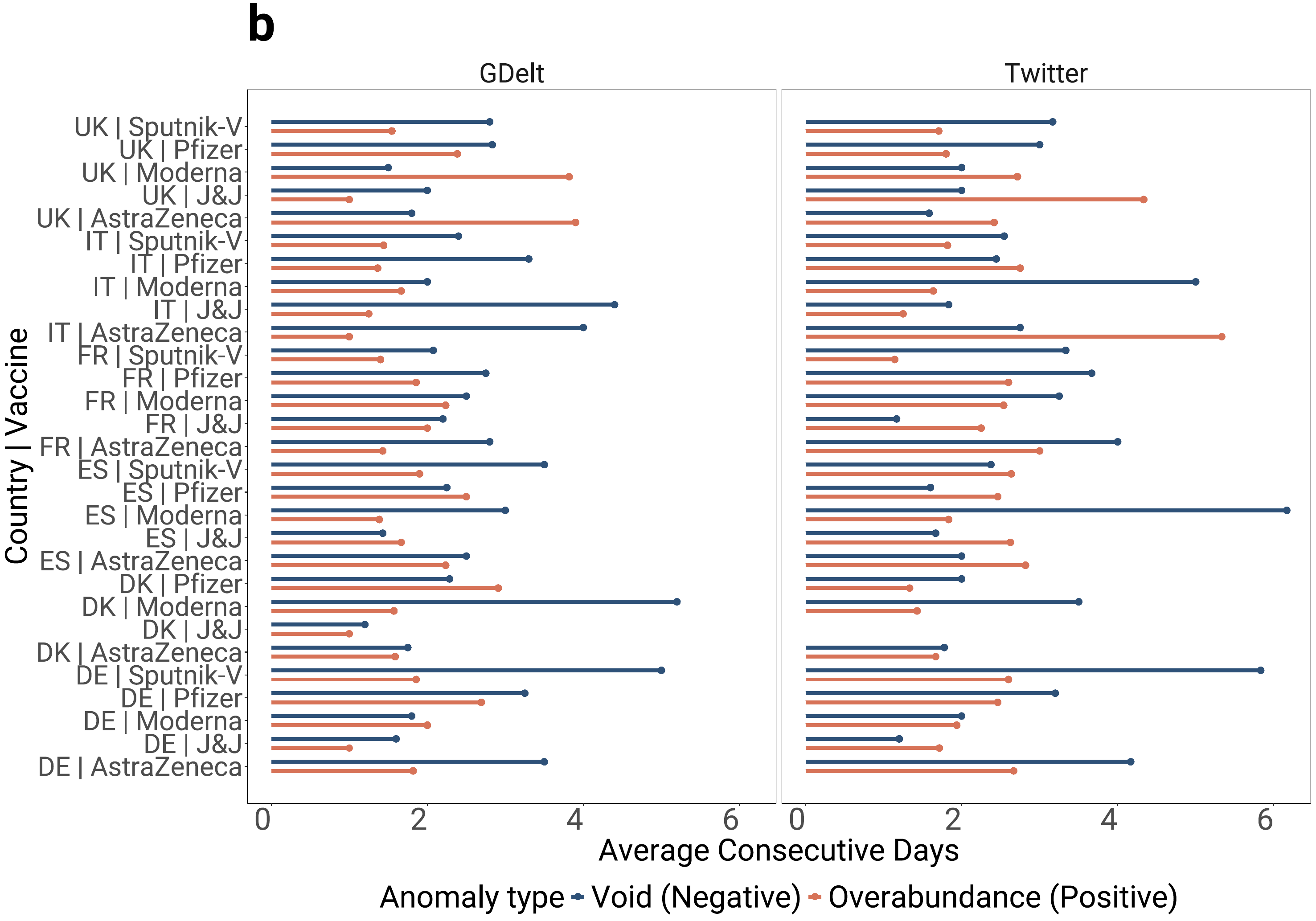}
    \caption{Average duration of Anomalies by Source Negative anomalies (blue) vs Positive anomalies (red)}
    \label{fig:wiki_cons}
\end{figure}

Beyond what presented in the main paper, we also analysed the distribution of anomalies before and after the start of the vaccination campaign (Figure~\ref{fig:gt_anomaly_det}a) using Google Trends, instead of Wikipedia, as the demand proxy. The results indicate that, prior to the vaccine rollout, the ecosystem remained largely stable, characterized by a predominance of non-anomalous behaviour. During this pre-rollout phase, both positive and negative anomalies were infrequent and weakly clustered, suggesting a general equilibrium between content production and information-seeking activity. This observation is consistent with the patterns observed when using Wikipedia pageviews as the demand proxy.
To further characterize the temporal dynamics of anomalies, we examined the average persistence of consecutive days classified as negative (information voids) or positive (information overabundance) anomalies (Figure~\ref{fig:gt_anomaly_det}b) using Google Trends. Persistence, defined as the number of consecutive daily anomalies sharing the same sign, reflects the duration of anomalous phases and indicates whether imbalances in the system are transient or sustained.
Figure~\ref{fig:wiki_cons} illustrates the average persistence of consecutive days classified as negative (information voids) or positive (information overabundance) anomalies for GDELT and Twitter, using Wikipedia pageviews as the demand proxy.

\section*{S8: Case studies}
As an additional analysis, we conduct a focused examination of several high-impact vaccine-related events that occurred during the study period. These case studies allow us to investigate in greater detail how highly debated events generate pronounced imbalances between information supply and demand. Specifically, we analyse: (a) the conditional authorisation of the Moderna vaccine by the European Medicines Agency (EMA); (b) reports by the Norwegian Medicines Agency concerning the investigation of deaths among elderly individuals who had received a COVID-19 vaccine \cite{wouters2022risk}, which were widely misinterpreted as evidence of vaccine-related fatalities; (c) the temporary suspension and subsequent reinstatement of the AstraZeneca vaccine following safety concerns; and (d) the recommendation by the CDC and FDA to pause the use of the Johnson \& Johnson COVID-19 vaccine due to reports of rare blood clots \cite{FDA2021JanssenPauseLifted}.

\begin{figure}[ht]
    \centering
    \includegraphics[width=1\linewidth]{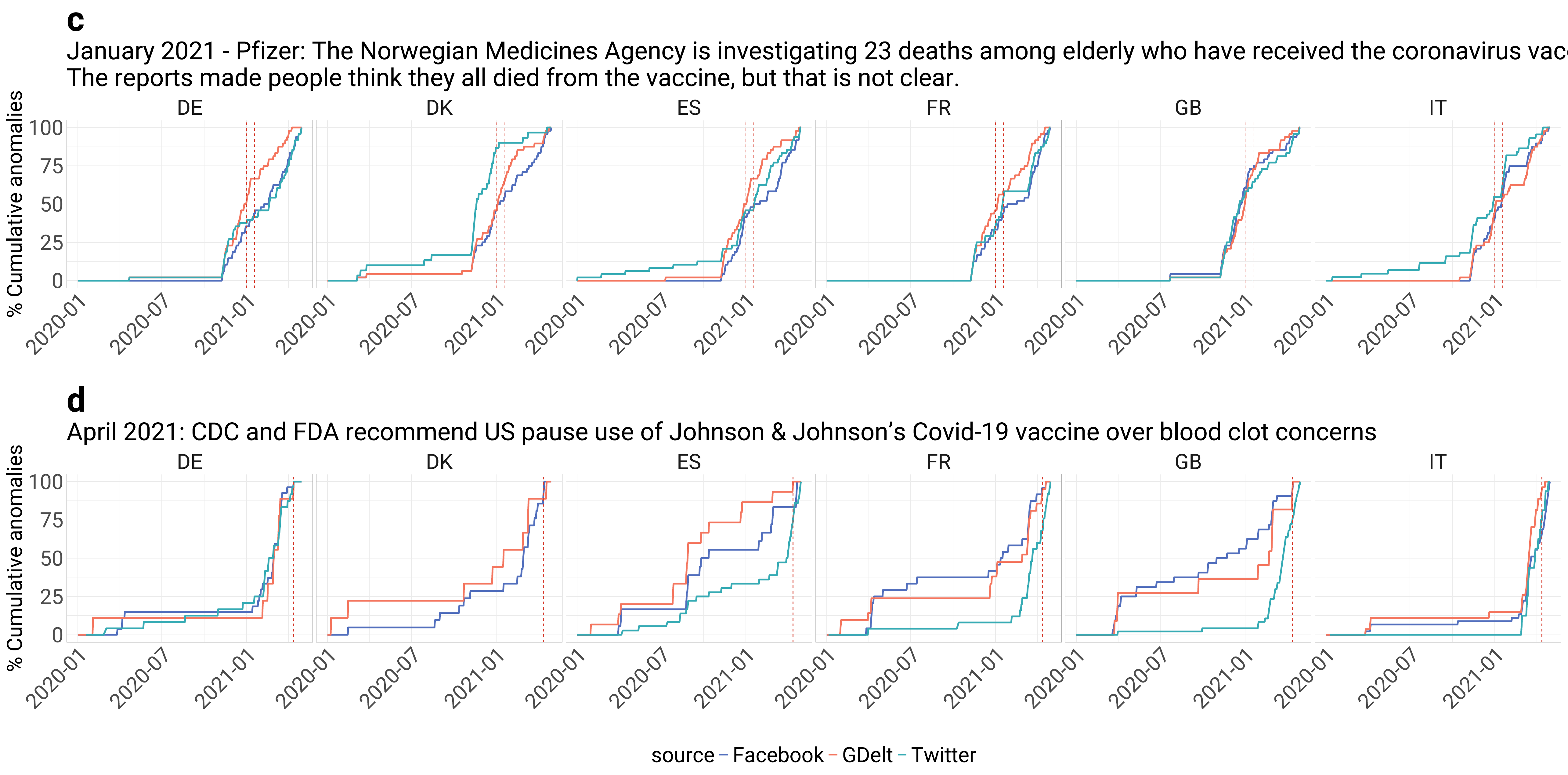}
    \caption{ (c) Cumulative anomalies associated with reports by the Norwegian Medicines Agency regarding the investigation of deaths among elderly individuals who had received a COVID-19 vaccine, which were widely misinterpreted as vaccine-related fatalities; red vertical lines indicate 1 January 2021, marking the onset of public concern, and 18 January 2021, corresponding to the official clarification by the Norwegian government. 
    (d) Cumulative anomalies related to the Johnson \& Johnson COVID-19 vaccine; the red vertical line marks the date on which the CDC and FDA recommended a pause in its use due to reports of rare blood clotting events (13 April 2021).}
    \label{fig:wiki_case}
\end{figure}

\begin{figure}[ht]
    \centering
    \includegraphics[width=1\linewidth]{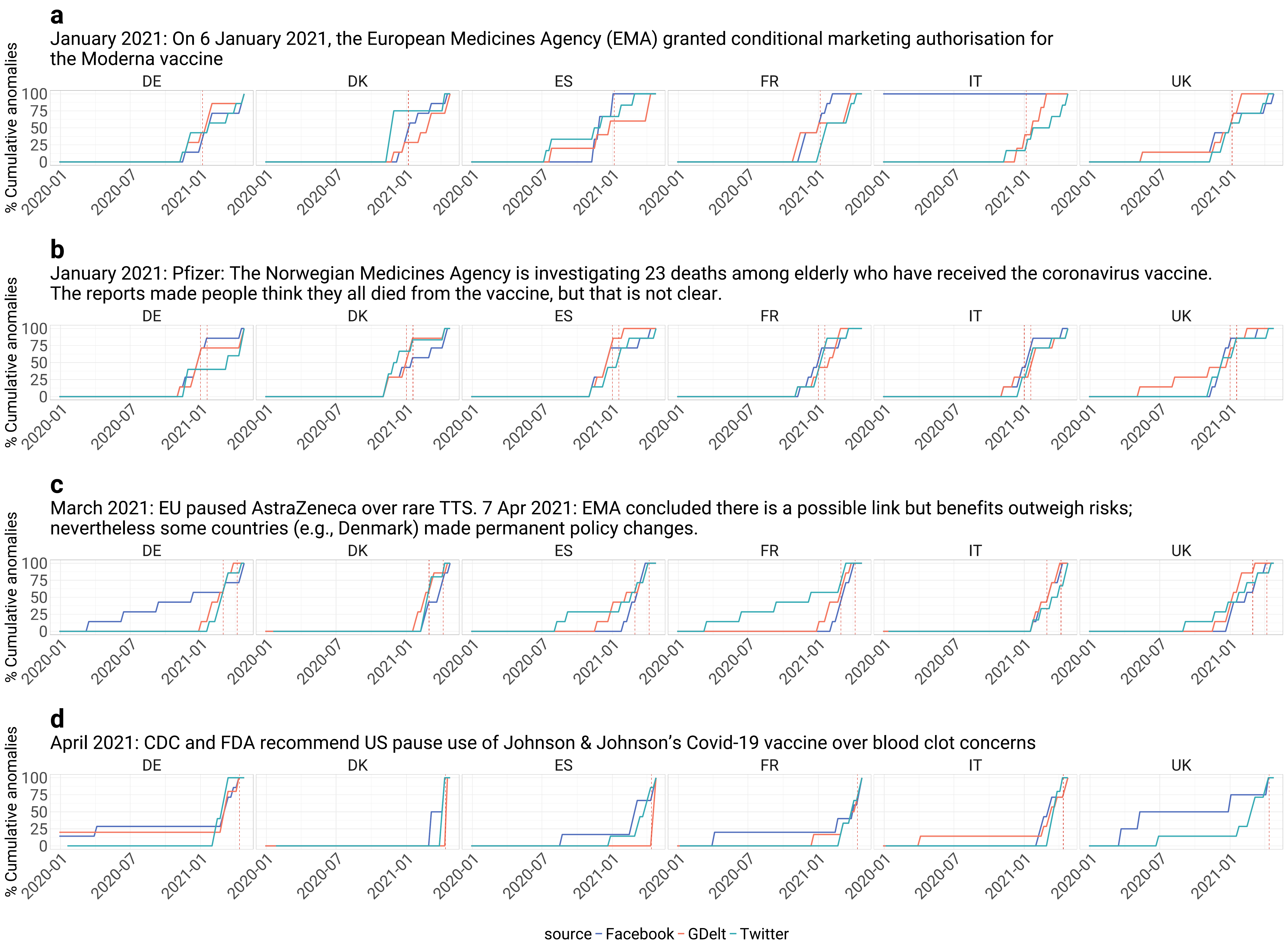}
    \caption{Case studies based on Google Trends. 
    (a) Cumulative anomalies for the Moderna COVID-19 vaccine by source and country, with the red line marking its EMA authorisation. (b) Cumulative anomalies linked to reports by the Norwegian Medicines Agency about deaths among vaccinated elderly individuals, with red lines indicating 1 January 2021 (onset of public concern) and 18 January 2021 (official government clarification). (c) Cumulative anomalies for the AstraZeneca COVID-19 vaccine by source and country, with red lines marking 1 March (rising TTS reports in Europe) and 7 April (EMA PRAC confirmation of a rare TTS link). (d) Cumulative anomalies for the Johnson \& Johnson COVID-19 vaccine, with the red line marking 13 April 2021, when the Centers for Disease Control and Prevention and Food and Drug Administration recommended a pause due to rare blood clotting events.}
    \label{fig:gt_case_SI}
\end{figure}

Figure~\ref{fig:wiki_case} shows the corresponding results for cases (b) and (d) using Wikipedia page views, while the remaining two case studies are discussed in the main text.
Figure~\ref{fig:gt_case_SI} presents the results based on Google Trends, used as a proxy for information demand. 

\section*{S9: Relationship between anomalies and misinformation prevalence}
We integrated the Facebook and Twitter datasets (using Wikipedia as the demand proxy) with NewsGuard data to obtain credibility scores for online sources. We then analysed how average credibility scores varied across periods of non-anomaly, void (negative anomaly), and overabundance (positive anomaly). To further investigate information quality under different anomaly conditions, we examined the distribution of posts across NewsGuard's credibility thresholds. 

\begin{figure}[ht]
    \centering
    \includegraphics[width=0.8\linewidth]{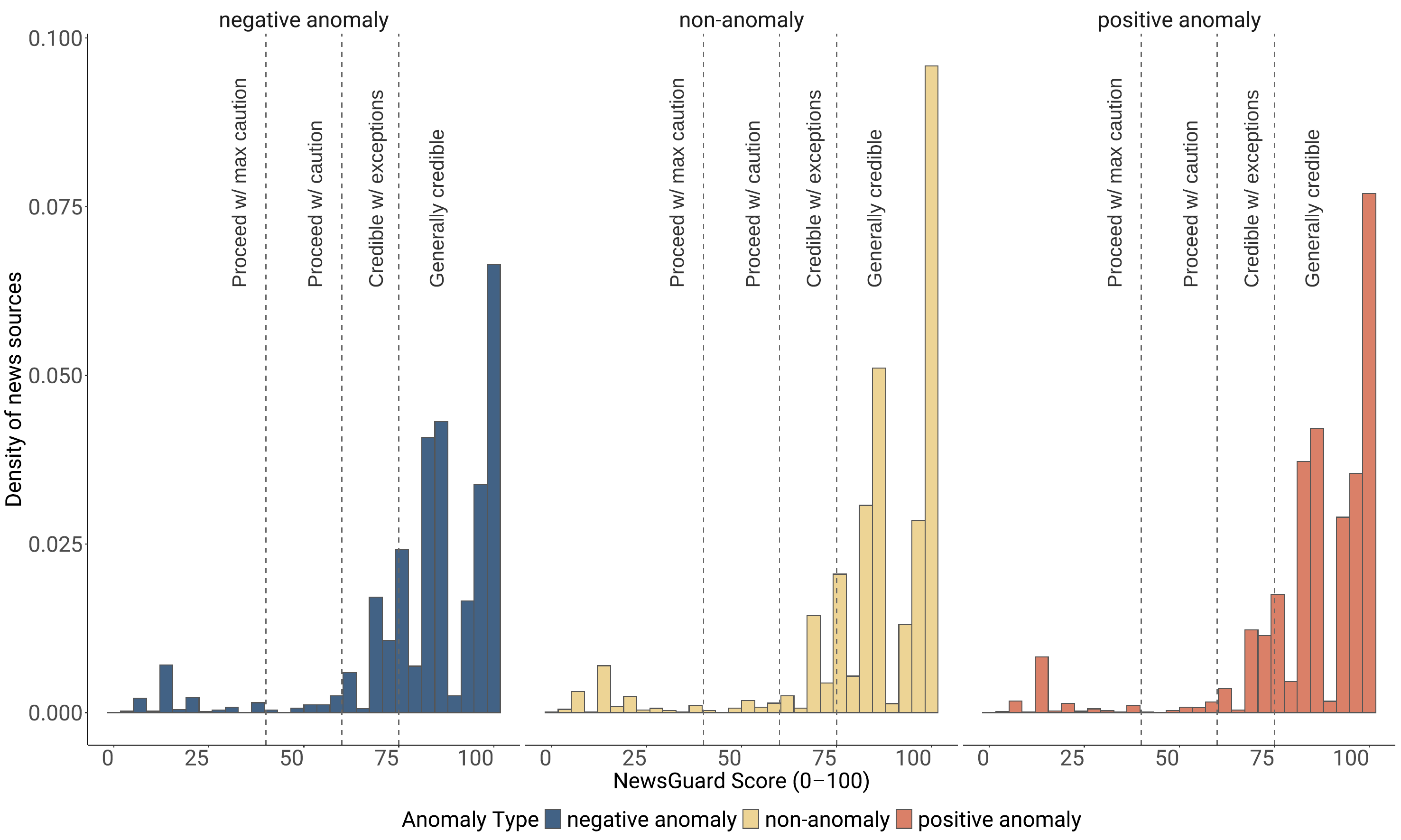}
    \caption{Distribution of Facebook posts by the NewsGuard-defined score.}
    \label{fig:fb_newsguard}
\end{figure}

\begin{figure}[H]
    \centering
    \includegraphics[width=0.8\linewidth]{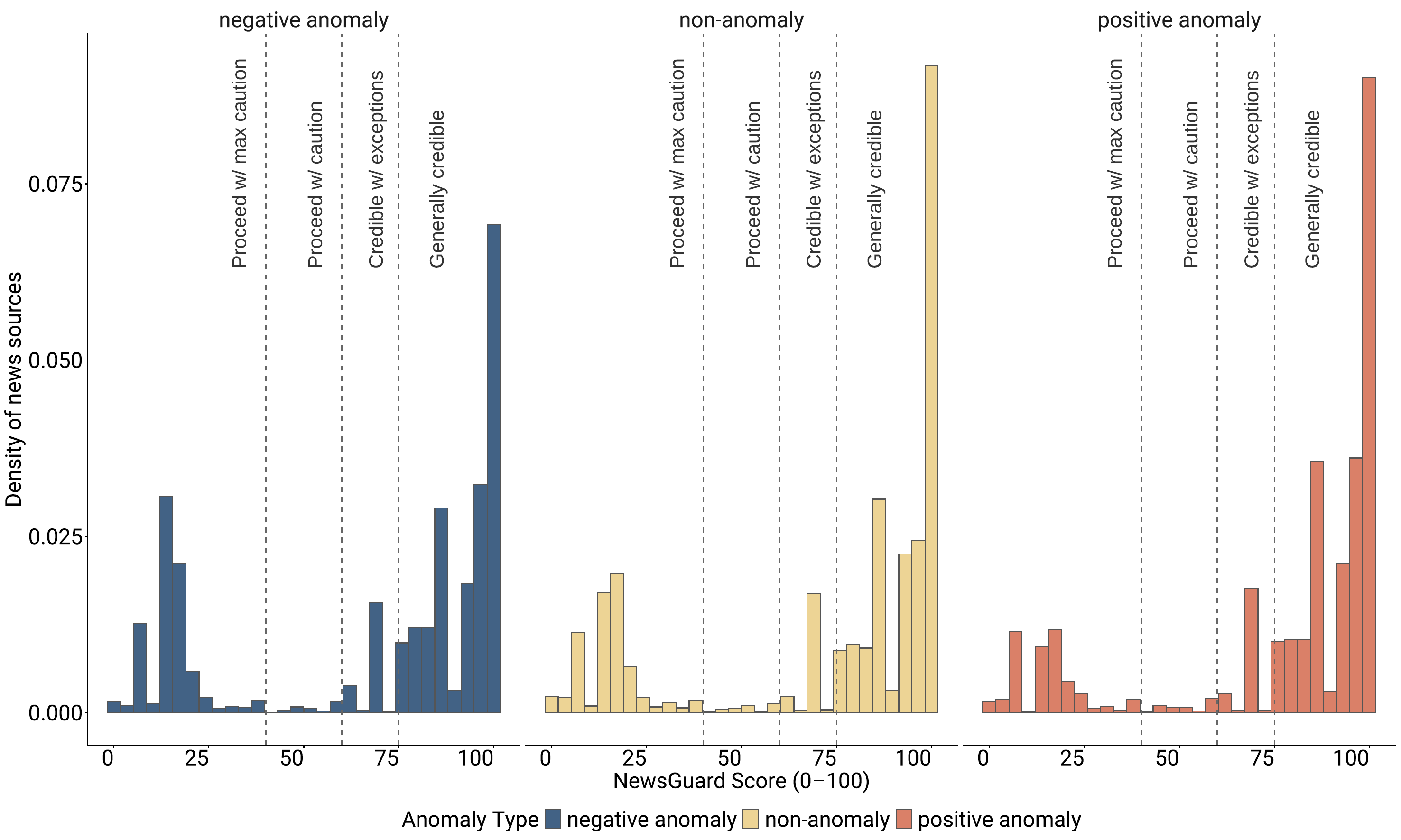}
    \caption{Distribution of Twitter posts by the NewsGuard-defined score}
    \label{fig:tw_newsguard}
\end{figure}

The resulting distributions are presented in Figure~\ref{fig:fb_newsguard} for Facebook and Figure~\ref{fig:tw_newsguard} for Twitter, with calculations performed from 8 December 2020 onward, corresponding to the launch of the COVID-19 vaccination campaign in the United Kingdom and the onset of heightened anomalous activity.

\end{document}